\documentclass[american,jfm]{revtex4-2}
\usepackage[T1]{fontenc}
\usepackage[latin9]{inputenc}
\usepackage{amsmath}
\usepackage{graphicx}
\usepackage{babel}
\begin{document}
\title{Solutions of the Navier-Stokes Equation Through Affine Transformations:
The Triad Triplet}
\author{{\"O}. D. G\"urcan, L. Manfredini and P. Morel}
\affiliation{Laboratoire de Physique des Plasmas, CNRS, Ecole Polytechnique, Sorbonne
Universit\'e, Universit\'e Paris-Saclay, Observatoire de Paris,
F-91120 Palaiseau, France}
\begin{abstract}
Considering triads in helical decomposition of three dimensional Navier-Stokes
turbulence, three consecutive triads of the same shape and class,
where the reference wave-number appears as the smallest, the middle
and the largest wave-numbers respectively, constitutes an interesting
object called the triad triplet that allows tracing the local transfer
in wave-number space due to a given shape and class of triads. It
is shown that, evolution equations for the triad triplet can be obtained
from the equations of the three legs of a single triad that are put
together via affine transformations involving scaling, rotation and
reflection. Known power law solutions corresponding to constant flux
of energy and helicity appear as exact solutions of a chain of triad
triplets, with particular implications for the triadic phases. More
interestingly, an exact, time dependent solution is available on an
isolated triad triplet, that can be expressed using what appears to
be straightforward generalization of Jacobi elliptic functions. These
solutions, constitute novel exact nonlinear solutions of the inviscid
limit of the Navier-Stokes equations.
\end{abstract}
\maketitle

\section{Introduction}

Navier-Stokes equations in the inviscid limit are invariant under
scaling and rotation. This suggests that its solutions $\mathbf{u}\left(\mathbf{x},t\right)$
would have these same invariances, at least locally. Invariance under
scaling might imply singularity, and in technical parlance, this is
expressed in terms of Hölder exponent, where one writes:
\begin{equation}
\left|\mathbf{u}\left(\mathbf{x}\right)-\mathbf{u}\left(\mathbf{y}\right)\right|\propto\left|\mathbf{x}-\mathbf{y}\right|^{H}\;\text{,}\label{eq:holder}
\end{equation}
which approaches a constant value as $\mathbf{y}\rightarrow\mathbf{x}$
instead of zero as one would have for regular smooth functions. Arguing
that such a solution for a given $H$ is a ``fractal'' with a corresponding
Hausdorff dimension $D\left(H\right)$, turbulence is then to be considered
as being made up of a set of such singularities belonging to different
fractals with different $D$'s gives us the usual multi-fractal picture
for turbulence\citep{benzi:84,frisch:85,frisch:91}. Noting that the
Fourier transform of $g\left(\ell\right)\equiv\left|\mathbf{x}-\mathbf{y}\right|^{H}$
gives $g_{k}\propto k^{-H-1}$ for the values of $H$ relevant for
three dimensional Navier-Stokes turbulence, the monofractal scaling
would therefore mean $\delta u_{k}\propto k^{-1/3}$. 

Since the interactions between wave-numbers occur via triadic interactions,
here we propose a formulation in terms of triads, where the fractal
structure can be identified in Fourier space as chains of triads that
are constructed from affine transformations of a single wave-number.
Using the helical decomposition of Waleffe\citep{waleffe:92,waleffe:93},
we recall that there are four different classes of triads depending
on the relative signs of the helicities of its larger and smaller
wave-numbers to the middle wave-number.

We argue that the minimal mono-fractal like object in $k$-space that
is actually \emph{self-affine} (i.e. that we can obtain by rotating,
scaling and reflecting the chiralities of an initial wave-number)
corresponds to what we call the triad triplet in this paper. We show
that the full time evolution of an isolated triad triplet can be solved
exactly when the amplitudes and phases are initialized to satisfy
well-known power law scalings. Considering only non-zero values for
the wave-numbers of a triad triplet, and computing the inverse Fourier
transforms, we can then obtain new, exact nonlinear solutions of the
Navier-Stokes equations, that can be expressed using a straightforward
generalization of the Jacobi elliptic functions\citep{abramowitz+stegun},
involving the inverse of an integral with multiple terms consisting
of two sine and one hyperbolic sine functions.

We also show that the Navier-Stokes equation, written in an infinite
domain using continuous Fourier transforms, can be reorganized into
a form where the $k$-space integrals can be written using triad triplets,
by considering the integrals over shape, orientation with respect
to the wave-number $\mathbf{k}$ and the sum over chirality classes.
This makes the cancellation in a power law solution transparent in
the equation directly.

We claim that the triad triplet is worth studying, in and of itself,
because i) it can be considered as a local (in Fourier space) diagnostic
tool, which allows us to focus on the role of a single shape and class
of triad in the overall cascade, and ii) it provides the basic nonlinear
oscillator that can eventually synchronize in order to result in a
burst of energy transfer in the sea of Fourier modes representing
turbulence. The role of such chains of nonlinear oscillators are found
to be essential in bursts of energy transfer in shell models\citep{manfredini:25b}.

The rest of the paper is organized as follows. In section \ref{sec:single_triad},
we recall the dynamics of a single triad and its phase evolution as
a background. In Section \ref{sec:triad_triplet} the concept of the
triad triplet is introduced, its basic evolution equation is written,
which is then used to re-write the Navier-Stokes system in a basis
consisting of triad-triplets, with integrals over shapes and orientations,
finally discussing evolution of triadic phases and conservation laws.
In Section \ref{sec:Exact-Solutions}, we first give two exact, constant
flux solutions corresponding to energy and helicity fluxes, on a triad
triplet chain, and then using the scaling of these solutions as initial
conditions, we find an exact time dependent solution, which is given
as a generalization of the Jacobi elliptic functions, which are also
compared with direct numerical solutions of the triad triplet equations.
In section \ref{sec:network}, we discuss how one can construct a
network of connected triads using triad triplets, and finally conclude
in section \ref{sec:Conclusion}.

\section{Dynamics of a Single Triad\label{sec:single_triad}}

Three dimensional Navier-Stokes turbulence can be formulated using
the helical decomposition of Waleffe\citep{waleffe:92,waleffe:93},
where a single triadic interaction with the ordering $p<k<q$, with
the helical modes $s_{k},s_{p},s_{q}$ can be written as:
\begin{align}
\partial_{t}u^{s_{k}}_{k} & =Q^{s_{k}s_{p}s_{q}}_{kpq}\left(s_{p}p-s_{q}q\right)u^{s_{p}*}_{p}u^{s_{q}*}_{q}\label{eq:t1}\\
\partial_{t}u^{s_{p}}_{p} & =Q^{s_{k}s_{p}s_{q}}_{kpq}\left(s_{q}q-s_{k}k\right)u^{s_{q}*}_{q}u^{s_{k}*}_{k}\label{eq:t2}\\
\partial_{t}u^{s_{q}}_{q} & =Q^{s_{k}s_{p}s_{q}}_{kpq}\left(s_{k}k-s_{p}p\right)u^{s_{k}*}_{k}u^{s_{p}*}_{p}\;\text{,}\label{eq:t3}
\end{align}
where 
\begin{equation}
Q^{s_{k}s_{p}s_{q}}_{kpq}=\frac{1}{4}s_{r}\sin\left(\alpha+\beta\right)\left(s_{k}+s_{p}\frac{p}{k}+s_{q}\frac{q}{k}\right)\;\text{,}\label{eq:q}
\end{equation}
with $s_{r}\equiv s_{k}s_{p}s_{q}$, and $\alpha$ and $\beta$ are
the angles between the middle and the smallest, and the largest and
the middle wave vectors of the triad respectively, and $Q^{s_{k}s_{p}s_{q}}_{kpq}$
is a geometric factor defining its shape/class. Note that even though
$\mathbf{k}$, $\mathbf{p}$ and $\mathbf{q}$ are three dimensional
wave-vectors, the triadic interaction only cares about their relative
orientations, which is determined by the relative magnitudes, $k$,
$p$ and $q$. In this sense, the triad above can be characterized
by the two angles $\alpha$ and $\beta$, which determine its shape
unambigously, and its helical mode class, where class 1 can be defined
as all the helical modes having the same sign (i.e. $s_{k}=s_{p}=s_{q}$)
and 2 to 4 can be defined as the largest (i.e. $s_{q}$) , the smallest
(i.e. $s_{p}$) and the middle leg (i.e. $s_{k}$) having a different
sign respectively {[}i.e. $\sigma_{g},\sigma_{h}=\left(+,+\right),\left(+,-\right),\left(-,+\right),\left(--\right)$
in that order, where $\sigma_{g}\equiv s_{k}s_{p}$, $\sigma_{h}\equiv s_{k}s_{q}${]}.

The four triad classes labeled as $1$ to $4$ (corresponding to classes
I, IV, II and III of Ref. \citealp{biferale:12} respectively), and
the instability assumption, which amounts to a linear stability calculation
with the initial energy in one of the legs of the triad, decides the
direction of the cascade. In general, for an isolated triad, the energy
goes from the energy containing mode (i.e. the pump) that has the
overall sign of its nonlinear interaction coefficient that is different
from the other two, towards the others. For simplicity, if the pump
mode is $k$, we call this triad ``$k$-unstable''. Sometimes we
will say that a triad is ``middle-unstable'', meaning that it is
unstable if the energy is initially in its middle leg. Considering
the form of the interaction coefficients, we note that the triads
that construct the triplet are middle-unstable if $\sigma_{h}>0$
and small-unstable if $\sigma_{h}<0$. In other words, no triad is
ever large-unstable, and it is the sign of $\sigma_{h}$ , which determines
the ``direction'' of the instability.

Since the instability assumption is in fact a manifestation of the
form of the triadic interaction equations, which are isomorphic to
Euler equations for a spinning top through a Poinsot construction,
extending the analogy, one can use the well known solution for this
system in terms of the Jacobi elliptic functions or more generally
in terms of the Weierstrass elliptic function (see Appendix \ref{sec:appA}). 

\subsection{Phase dynamics:}

We can obtain the phase equations for a single triad by substituting
$u^{s_{i}}_{i}=U^{s_{i}}_{i}e^{i\theta^{s_{i}}_{i}}$, where $i=\left\{ k,p,q\right\} $
in Eqn. (\ref{eq:t1}-\ref{eq:t3}), multiplying by $e^{-i\theta^{s_{i}}_{i}}$
and taking the imaginary part. For example for $k$, one obtains:
\[
\partial_{t}\theta^{s_{k}}_{k}=-Q\left(s_{p}p-s_{q}q\right)\frac{U^{s_{p}}_{p}U^{s_{q}}_{q}}{U^{s_{k}}_{k}}\sin\phi^{s_{k}s_{p}s_{q}}_{kpq}
\]
where $\phi^{s_{k}s_{p}s_{q}}_{kpq}=\theta^{s_{k}}_{k}+\theta^{s_{p}}_{p}+\theta^{s_{q}}_{q}$
is the triadic phase and $Q$ is a shortcut for $Q^{s_{k}s_{p}s_{q}}_{kpq}$.
We can then obtain the triadic phase equations by combining the individual
phase equations. However a more direct way is to consider the equations
for the triple product $\chi^{s_{k}s_{p}s_{q}}_{kpq}\equiv u^{s_{k}}_{k}u^{s_{p}}_{p}u^{s_{q}}_{q}$,
which can be obtained by multiplying the equations (\ref{eq:t1}-\ref{eq:t3})
by $u^{s_{p}}_{p}u^{s_{q}}_{q}$, $u^{s_{q}}_{q}u^{s_{k}}_{k}$ and
$u^{s_{k}}_{k}u^{s_{p}}_{p}$ respectively and the summing the results:
\begin{align}
\partial_{t}\chi^{s_{k}s_{p}s_{q}}_{kpq}=Q\bigg[ & \left(s_{p}p-s_{q}q\right)U^{s_{p}2}_{p}U^{s_{q}2}_{q}+\left(s_{q}q-s_{k}k\right)U^{s_{q}2}_{q}U^{s_{k}2}_{k}\nonumber \\
 & +\left(s_{k}k-s_{p}p\right)U^{s_{k}2}_{k}U^{s_{p}2}_{p}\bigg]\quad\text{.}\label{eq:tpeq}
\end{align}
Substituting $\chi^{s_{k}s_{p}s_{q}}_{kpq}=\left|\chi^{s_{k}s_{p}s_{q}}_{kpq}\right|e^{i\phi^{s_{k}s_{p}s_{q}}_{kpq}}$
in (\ref{eq:tpeq}), multiplying it by $e^{-i\phi^{s_{k}s_{p}s_{q}}_{kpq}}$
and taking the imaginary part gives the triadic phase equation:
\begin{align}
\partial_{t}\phi^{s_{k}s_{p}s_{q}}_{kpq}=-Q\bigg[ & \left(s_{p}p-s_{q}q\right)U^{s_{p}}_{p}U^{s_{q}}_{q}/U^{s_{k}}_{k}+\left(s_{q}q-s_{k}k\right)U^{s_{q}}_{q}U^{s_{k}}_{k}/U^{s_{p}}_{p}\nonumber \\
 & +\left(s_{k}k-s_{p}p\right)U^{s_{k}}_{k}U^{s_{p}}_{p}/U^{s_{k}}_{q}\bigg]\sin\phi^{s_{k}s_{p}s_{q}}_{kpq}\label{eq:phpeq}
\end{align}
Note that the right hand side vanishes for $\phi^{s_{k}s_{p}s_{q}}_{kpq}=0$
or $\pi$, or when the inside of the square bracket vanishes as in
the case of power law solutions to be discussed below. These are in
a sense the ``fixed points'' of this equation, but they come either
from the triadic phases being $0$ or $\pi$, regardless of the amplitudes,
or the amplitudes satisfying the power law solutions exactly, regardless
of the phases, any deviation from either results in nonlinear perturbations
with feedback loops. 

Another interesting observation is that since the right hand side
of (\ref{eq:tpeq}) is real, by taking the imaginary part of (\ref{eq:tpeq})
directly we can write:
\begin{equation}
\partial_{t}\left(\left|\chi^{s_{k}s_{p}s_{q}}_{kpq}\right|\sin\phi^{s_{k}s_{p}s_{q}}_{kpq}\right)=0\;\text{,}\label{eq:imphzero}
\end{equation}
which can be solved exactly as:
\begin{equation}
\sin\phi^{s_{k}s_{p}s_{q}}_{kpq}=\frac{M}{U^{s_{k}}_{k}U^{s_{p}}_{p}U^{s_{q}}_{q}}\label{eq:sinphi}
\end{equation}
where $M$ is a constant related to initial conditions and the Hamiltonian
of the system.

\section{Dynamics of a Triad Triplet:\label{sec:triad_triplet}}

Considering three consecutive triads of the same shape and class,
where the reference wave-number $k$ is the smallest, the middle and
the largest wave-number respectively, gives us what we call the triad
triplet. Such a construction is interesting because it is capable
of describing a ``local'' flux through $k$-space for a particular
shape (and for the helical decomposition, also a particular class)
of triads. The general equation of the triad triplet can be obtained
by scaling and rotating Eqns. (\ref{eq:t2}) and (\ref{eq:t3}), which
are written in such a way that they represent the smaller and larger
legs of the initial triad, for which $k$ was the middle wave number.
This allows us to write down the equations for $k$ where $k$ is
the larger or the smaller wave-number of the triad, keeping the ratios
between $k/p$ and $k/q$ unchanged, so that the shape of the triad
(the angles between the consecutive wavenumbers) do not change. For
the helical decomposition, we also need to transform the $s_{i}$'s
in such a way that we obtain $s_{k}$ and $k$ without changing the
triad class. 

For Eqn. (\ref{eq:t2}) this can be achieved by multiplying all the
wavenumbers by $k/p$ and all the $s_{i}'s$ by $s_{k}/s_{p}$ (or
equivalently $s_{k}s_{p}$), which defines an affine transformation
(scaling, rotation and reflection) that gives:
\begin{equation}
\partial_{t}u^{s_{k}}_{k}=\frac{k^{2}}{p}Q^{s_{k}s_{p}s_{q}}_{kpq}\left(s_{r}\frac{q}{k}-s_{p}\right)u^{s_{r}*}_{\frac{qk}{p}}u^{s_{p}*}_{\frac{k^{2}}{p}}\;\text{,}\label{eq:c1}
\end{equation}
where $s_{r}=s_{k}s_{p}s_{q}$ as before, and $Q^{s_{k}s_{p}s_{q}}_{kpq}$
can be written as a function only of the ratios $p/k$, $q/k$, $s_{p}/s_{k}$
and $s_{q}/s_{k}$ and since those do not change, $Q$ acts as a geometric
factor, and it remains invariant under our affine transformations. 

Similary we can obtain the equation where $k,s_{k}$ is the largest
leg of the same class of triads by multiplying all the wavenumbers
of Eqn. (\ref{eq:t3}), by $k/q$ and all the $s_{i}'s$ by $s_{k}/s_{q}$
:
\begin{equation}
\partial_{t}u^{s_{k}}_{k}=\frac{k^{2}}{q}Q^{s_{k}s_{p}s_{q}}_{kpq}\left(s_{q}-s_{r}\frac{p}{k}\right)u^{s_{q}*}_{\frac{k^{2}}{q}}u^{s_{r}*}_{\frac{pk}{q}}\;\text{,}\label{eq:c2}
\end{equation}
Putting it all together, we can write the equation for $u^{s_{k}}_{k}$
with contributions from these three consecutive triads of the same
shape and class as:
\begin{align}
\partial_{t}u^{s_{k}}_{k}=kQ^{s_{k}s_{p}s_{q}}_{kpq}\bigg[ & \frac{k}{q}\left(s_{q}-s_{r}\frac{p}{k}\right)u^{s_{q}*}_{\frac{k^{2}}{q}}u^{s_{r}*}_{\frac{pk}{q}}+\left(s_{p}\frac{p}{k}-s_{q}\frac{q}{k}\right)u^{s_{p}*}_{p}u^{s_{q}*}_{q}\nonumber \\
 & +\frac{k}{p}\left(s_{r}\frac{q}{k}-s_{p}\right)u^{s_{r}*}_{\frac{qk}{p}}u^{s_{p}*}_{\frac{k^{2}}{p}}\bigg]\;\text{.}\label{eq:tt}
\end{align}
Eqn. (\ref{eq:tt}) is our key result, and it describes the evolution
of the Fourier mode $u^{s_{k}}_{k}$ of a reference wave-number $\mathbf{k}$,
connected to a set of three consecutive triads of the same shape and
class, such that the reference wave-number is the smallest, the middle
and the largest leg of each of these triads respectively. 

Integrating this over the azimuthal angle around $\mathbf{k}$ (i.e.
any rotation of $\mathbf{p}$ and $\mathbf{q}$ around the axis defined
by $\hat{\mathbf{k}}$, would give us a different orientation of the
triplet with the same geometric factor), and the shape of the triad
(for example defined by the magnitudes of $p$ and $q$ or by the
angles $\alpha$ and $\beta$, which would of course change the geometric
factor), and finally considering the four different classes, one can
compute the full convolution sum/integral. However it should be noted
that, on a discrete grid, some of those triads would not exist, and
therefore the perfect self-similarity that is implicit in these symmetry
transformations would not be respected on a regular wave-number space
grid. Nonetheless, the concept of the triad triplet is useful in tracing
the cascade through similar kinds of triads accross the Fourier space,
especially deep in the inertial range. It is also probably useful
for studying inviscid singular solutions in $k$-space. The formulation
could potentially be extended to the discrete case using recurrence
relations corresponding to these affine transformations as is done
in shell models\citep{manfredini:25}. 

Note that we can also define $k/p=g$ and $q/k=h$, $\sigma_{g}\equiv s_{p}s_{k}$
and $\sigma_{h}=s_{q}s_{k}$ in order to write everything in terms
of $k$, $h$ and $g$, as:
\begin{align}
\partial_{t}u^{s_{k}}_{k}=s_{k}kQ\left(g,h\right)^{\sigma_{g}\sigma_{h}}\bigg[ & h^{-1}\sigma_{h}\left(1-\sigma_{g}g^{-1}\right)u^{s_{q}*}_{kh^{-1}}u^{s_{r}*}_{kh^{-1}g^{-1}}+\left(\sigma_{g}g^{-1}-\sigma_{h}h\right)u^{s_{p}*}_{kg^{-1}}u^{s_{q}*}_{kh}\nonumber \\
 & +g\sigma_{g}\left(\sigma_{h}h-1\right)u^{s_{r}*}_{kgh}u^{s_{p}*}_{kg}\bigg]\;\text{,}\label{eq:tthg}
\end{align}
which makes the connection to shell models\citep{biferale:03} (with
$g^{n}=h^{m}$ where $n,m$ are integers) and other self-similar cascade
models including spiral chains\citep{gurcan:19} more transparent.
Note that $Q$ is now a function only of the ratios $g$ and $h$
and the relative signs $\sigma_{g}$ and $\sigma_{h}$. The angles
$\alpha$ and $\beta$ can be obtained from the triad condition, which
can be written using $g$ and $h$ as
\begin{equation}
1+g^{-1}e^{-i\alpha}+he^{i\beta}=0\label{eq:gsum}
\end{equation}
which gives:
\begin{align}
\alpha & =\arccos\left(\frac{h^{2}-g^{-2}-1}{2g^{-1}}\right)=\arccos\left(\frac{q^{2}-k^{2}-p^{2}}{2kp}\right)\label{eq:alp1}\\
\beta & =\arccos\left(\frac{g^{-2}-h^{2}-1}{2h}\right)=\arccos\left(\frac{p^{2}-k^{2}-q^{2}}{2qk}\right)\label{eq:bet1}
\end{align}
or
\begin{equation}
g=-\frac{\sin\left(\alpha+\beta\right)}{\sin\beta}\;\text{,}\quad h=-\frac{\sin\alpha}{\sin\left(\alpha+\beta\right)}\label{eq:gh_ab}
\end{equation}
and $s_{p},p$ and $s_{q},q$ can be obtained from $s_{k},k$ by scaling
by $g^{-1}$ and $h$ and rotating by $-\alpha$ and $\beta$ and
reflecting by $\sigma_{g}$ and $\sigma_{h}$ respectively. Note that
in this sense, the initial triad itself can also be constructed through
a pair of affine transformations of a single wave-vector $\mathbf{k}$.
The same pair of transformations can then be used on the triad, in
order to construct the triad triplet. Using the same transformation
we can keep expanding a network of triads that can span the $k$-space.
Note that from Eqns. (\ref{eq:alp1}) and (\ref{eq:bet1}), we must
have $g^{-1}\geq h-1$ .

\begin{figure}
\begin{centering}
\includegraphics[width=0.98\textwidth]{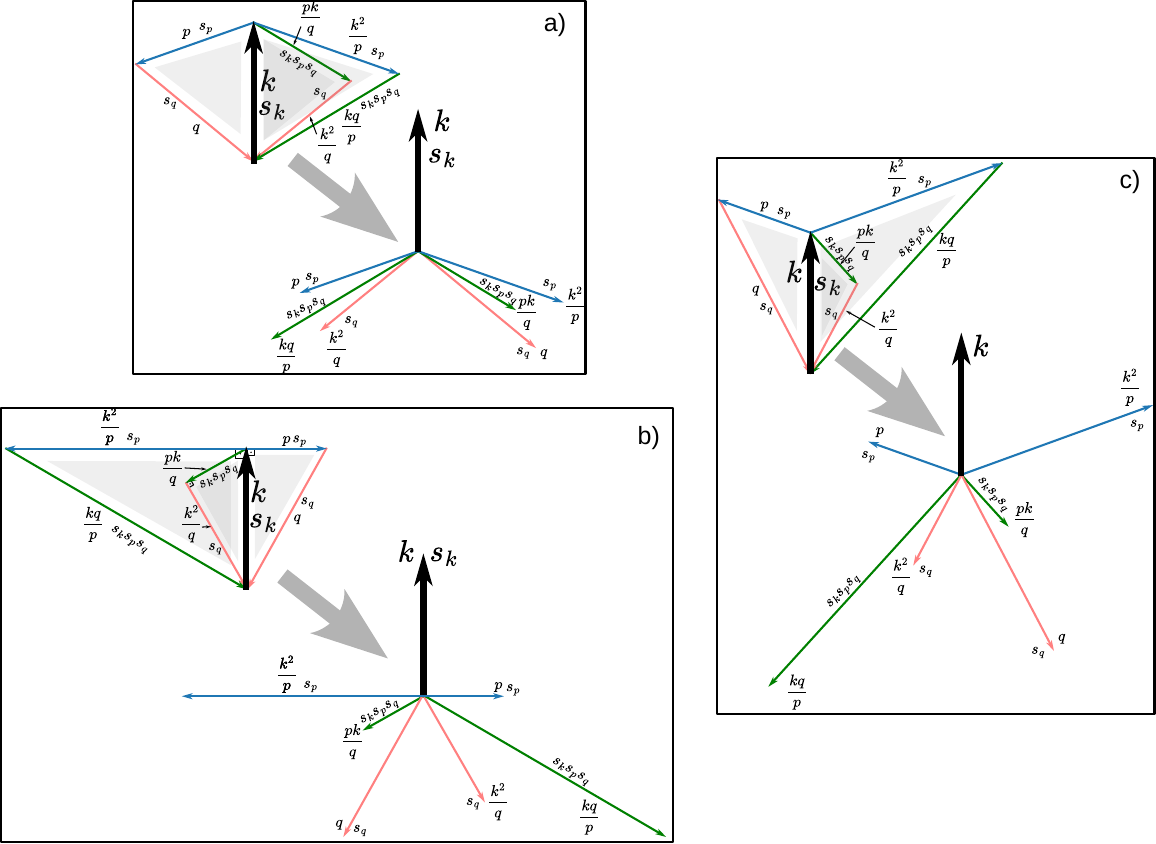}
\par\end{centering}
\caption{\label{fig:types_tt}Examples of triad triplets, showing both the
triads themselves and the wave-vectors, where the wave-vectors with
helical modes of signs $s_{k}$, $s_{p}$, $s_{q}$ and $s_{k}s_{p}s_{q}$,
are shown in black, blue, red and green respectively (if in color).
Top left figure labeled as a) shows an acute or local triplet consisting
of acute triangles, which describes local interactions bottom left
figure labeled as b) shows a right triplet, since it consists of three
right triangles, and the figure on the right labelled as c) shows
an obtuse or elongated triplet consisting of three obtuse triangles
consisting of more nonlocal interactions. The ratio of the smallest
to largest wave-numbers of a triad triplet (i.e. the two green legs
in the above figures) can be used as a measure of its nonlocality.
We see that for the examples above, both the obtuse and the right
triplets are actually similarly elongated. Note also that when we
take the local limit, we only consider the interactions of the type
shown in the top left figure.}
\end{figure}

Figure \ref{fig:types_tt} shows three different types of triad triplets,
which can be denoted as acute, right and obtuse, with acute triplets
describing mainly local interactions while right or obtuse triplets,
describing more nonlocal interactions depending on their elongation,
which, could be defined as the ratio of the largest to the smallest
leg i.e. $q/p=gh$.

\subsection{Navier Stokes in triad triplet basis\label{subsec:nsintt}}

The Navier stokes equation can be written, using continuous Fourier
transforms and helical decomposition as:
\begin{equation}
\partial_{t}u^{s_{k}}\left(\mathbf{k}\right)+\sum_{s_{p},s_{q}}\iint Q^{s_{k}s_{p}s_{q}}_{kpq}\left(s_{p}p-s_{q}q\right)u^{s_{p}*}\left(\mathbf{p}\right)u^{s_{q}*}\left(\mathbf{q}\right)\delta^{3}\left(\mathbf{k}+\mathbf{p}+\mathbf{q}\right)d^{3}\mathbf{p}d^{3}\mathbf{q}=-\nu k^{2}u^{s_{k}}\left(\mathbf{k}\right)\label{eq:nsint}
\end{equation}
where the integral with the dirac delta function can be re-interpreted
as being over different triad shapes and sizes, and the sum over $s_{p}$
and $s_{q}$ is equivalent to a sum over different classes. 

Dividing the range of integration into three different regions $p<k<q$,
$k<q'<p'$ and $q''<p''<k$, and parametrizing the triad in each region
using $g$ and $h$ (see Appendix \ref{sec:appC}) we can write:
\begin{align}
\partial_{t}u^{s_{k}}\left(\mathbf{k}\right) & +s_{k}k^{4}\int^{\infty}_{1}\int^{\infty}_{1}\int^{2\pi}_{0}\sum_{\sigma_{p},\sigma_{q}}Q\left(g,h\right)^{\sigma_{g}\sigma_{h}}\bigg[\nonumber \\
 & +h^{-7}\sigma_{h}\left(1-\sigma_{g}g^{-1}\right)u^{s_{q}*}\left(kh^{-1},-\beta,\phi\right)u^{s_{r}*}\left(kh^{-1}g^{-1},-\alpha-\beta,\phi\right)\nonumber \\
 & +\left(\sigma_{g}g^{-1}-\sigma_{h}h\right)u^{s_{p}*}\left(kg^{-1},-\alpha,\phi\right)u^{s_{q}*}\left(kh,\beta,\phi\right)\\
 & +g^{7}\sigma_{g}\left(\sigma_{h}h-1\right)u^{s_{r}*}\left(kgh,\alpha+\beta,\phi\right)u^{s_{p}*}\left(kg,\alpha,\phi\right)\nonumber \\
 & \bigg]g^{-3}hdgdhd\phi\nonumber \\
 & =-\nu k^{2}u^{s_{k}}\left(\mathbf{k}\right)\label{eq:nstt}
\end{align}
where $\alpha$ and $\beta$ are defined through Eqns. \ref{eq:alp1}
and \ref{eq:bet1} in terms of $g$ and $h$. Note that this is a
simple rewrite of the Navier-Stokes equation, in terms of triad triplets,
where the integral is now over shapes, that are parametrized by $g$
and $h$ and the sum is over classes parametrized by $\sigma_{g}$
and $\sigma_{h}$, with an additional integral over orientations $\phi$,
defined with respect to the direction of $\mathbf{k}$. The variables
$u^{s_{k'}}\left(k',\alpha,\phi\right)$ scale as $u^{s_{k'}}_{k'}/dk'^{3}$,
which explains the additional $h^{-6}$ and $g^{6}$ factors in the
first and the last terms in comparison to Eqn. \ref{eq:tthg}.

This form allows us to see that, if we tailored a single fractal solution
with a single triad shape and a power law with the correct phase relations
it would actually satisfy the full Navier-Stokes equation, because
of the term by term cancellation in Eqn. (\ref{eq:nstt}).

Note that the geometric factor $Q$ is oblivious to triad orientation
or scale. As a result the triad triplet equation, i.e. Eqns. \ref{eq:tt}-\ref{eq:tthg},
does not care about orientation either. We can see this also in Eqn.
\ref{eq:nstt}, where the only dependence on the orientation $\phi$
comes from the dependence of the Fourier coefficients $u^{s_{k}}\left(k,\theta,\phi\right)$
themselves. This means that the triad triplet equations do not have
to involve only the wave-numbers that lie on a plane. Alternatively,
the exact, single fractal solutions that we discuss below, that rely
on a particular choice of the triadic phases, may actually involve
different orientations at different levels. 

In other words, if the difference between different orientations of
a triad with the same shape, is only in their phases (as would be
in a power law solution), the flux would flow through those triads
whose orientations satisfy the correct phase relation. Furthermore,
since the $\mathbf{k}$ in Eqn. \ref{eq:tt} is a vector, while a
rotation around itself (i.e. $\phi$), doesn't do anything, we can
consider rotations by $\theta$ and $\phi$ which can be defined with
respect to a fixed Cartesian coordinate system to compute energy flux
at a given scale assuming isotropy.

\subsection{Triad phase dynamics:}

Going back to the discrete form, the equation for the triadic correlations,
equivalent to Eqn. (\ref{eq:tpeq}) for a single triad, takes the
form:
\begin{align}
\partial_{t}\chi^{s_{k}s_{p}s_{q}}_{kpq}= & Q^{s_{k}s_{p}s_{q}}_{kpq}\bigg[\left(\frac{\left(s_{p}p-s_{q}q\right)}{E^{s_{k}}_{k}}+\frac{\left(s_{q}q-s_{k}k\right)}{E^{s_{p}}_{p}}+\frac{\left(s_{k}k-s_{p}p\right)}{E^{s_{q}}_{q}}\right)\left|\chi^{s_{k}s_{p}s_{q}}_{kpq}\right|^{2}\nonumber \\
 & +\frac{k\left(s_{q}k-s_{r}p\right)}{qE^{s_{k}}_{k}}\chi^{s_{q}s_{r}s_{k}*}_{kh^{-1},ph^{-1},qh^{-1}}\chi^{s_{k}s_{p}s_{q}}_{kpq}+\frac{q\left(s_{r}q-s_{p}k\right)}{pE^{s_{k}}_{k}}\chi^{s_{p}s_{k}s_{r}*}_{kg,pg,qg}\chi^{s_{k}s_{p}s_{q}}_{kpq}\bigg]\;\text{.}\label{eq:tttc}
\end{align}
Substituting $\chi^{s_{k}s_{p}s_{q}}_{kpq}=\left|\chi^{s_{k}s_{p}s_{q}}_{kpq}\right|e^{i\phi^{s_{k}s_{p}s_{q}}_{kpq}}$
, multiplying by $\left|\chi^{s_{k}s_{p}s_{q}}_{kpq}\right|^{-1}e^{-i\phi^{s_{k}s_{p}s_{q}}_{kpq}}$
and taking the imaginary part gives the triadic phase equation:
\begin{align}
\partial_{t}\phi^{s_{k}s_{p}s_{q}}_{kpq}= & -Q^{s_{k}s_{p}s_{q}}_{kpq}\bigg[\left(\frac{\left(s_{p}p-s_{q}q\right)}{E^{s_{k}}_{k}}+\frac{\left(s_{q}q-s_{k}k\right)}{E^{s_{p}}_{p}}+\frac{\left(s_{k}k-s_{p}p\right)}{E^{s_{q}}_{q}}\right)\left|\chi^{s_{k}s_{p}s_{q}}_{kpq}\right|\sin\phi^{s_{k}s_{p}s_{q}}_{kpq}\nonumber \\
 & +\frac{k\left(s_{q}k-s_{r}p\right)}{qE^{s_{k}}_{k}}\left|\chi^{s_{q}s_{r}s_{k}}_{kh^{-1},ph^{-1},qh^{-1}}\right|\sin\phi^{s_{q}s_{r}s_{k}}_{kh^{-1},ph^{-1},qh^{-1}}+\frac{k\left(s_{r}q-s_{p}k\right)}{pE^{s_{k}}_{k}}\left|\chi^{s_{p}s_{k}s_{r}}_{kg,pg,qg}\right|\sin\phi^{s_{p}s_{k}s_{r}*}_{kg,pg,qg}\bigg]\;\text{,}\label{eq:phase}
\end{align}
where the phase of a particular triad, of a given shape (defined by
$g$ and $h$), and class (defined by $\sigma_{g}$ and $\sigma_{h}$),
is determined by a combination of a self interaction term and two
terms representing the couplings to smaller and larger triads of the
triplet. Note that the self interaction term is the same as the one
for a single triad, given in Eqn. \ref{eq:phpeq}, and that the standard
equipartition solution, which corresponds to $E_{k}\propto\text{const}.$
here, makes its coefficient vanish.

\subsection{Conservation laws:}

Using Eqn. (\ref{eq:tthg}) we can write the equation for energy and
helicity conservation for a single triad triplet as:
\begin{align*}
\partial_{t}E^{s_{k}}_{k}= & T^{s_{k}s_{p}s_{q}}_{k,p,q}+T^{s_{k}s_{q}s_{r}}_{qh^{-1},kh^{-1},ph^{-1}}+T^{s_{k}s_{r}s_{p}}_{pg,qg,kg}\\
= & T^{s_{k}}_{k}\left(g^{-1},h;\sigma_{g},\sigma_{h}\right)+T^{s_{p}}_{kg}\left(1,gh;1,\sigma_{h}\sigma_{g}\right)+T^{s_{q}}_{kh^{-1}}\left(g^{-1}h^{-1},1;\sigma_{g}\sigma_{h},1\right)
\end{align*}
where:
\begin{equation}
T^{s_{k}s_{p}s_{q}}_{k,p,q}\equiv\text{Re}\left[Q\left(s_{p}p-s_{q}q\right)\chi^{s_{k}s_{p}s_{q}}_{kpq}\right]\label{eq:Ttt-1}
\end{equation}
is the energy transfer from the pair $\left(\mathbf{p},s_{p}\right)$
and $\left(\mathbf{q},s_{q}\right)$ to the reference node $\left(\mathbf{k},s_{k}\right)$
and $Q=Q\left(g,h\right)^{\sigma_{g}\sigma_{h}}$ is the geometric
factor as before. Equivalently
\[
T^{s_{k}}_{k}\left(g^{-1},h;\sigma_{g},\sigma_{h}\right)=s_{k}k\text{Re}\left[Q\left(\sigma_{g}g^{-1}-\sigma_{h}h\right)\chi^{s_{k}}_{k}\left(g^{-1},h;\sigma_{g}\sigma_{h}\right)\right]
\]
with
\[
\chi^{s_{k}}_{k}\left(g^{-1},h;\sigma_{g}\sigma_{h}\right)=\chi^{s_{k}s_{p}s_{q}}_{kpq}=u^{s_{k}}_{k}u^{s_{k}\sigma_{g}}_{kg^{-1}}u^{s_{k}\sigma_{h}}_{kh}\;\text{.}
\]
Similarly for the helicity, we can write:
\begin{align*}
\partial_{t}H^{s_{k}}_{k}= & T^{\left(H\right)s_{k}s_{p}s_{q}}_{kpq}+T^{\left(H\right)s_{k}s_{r}s_{p}}_{pg,qg,kg}+T^{\left(H\right)s_{k}s_{q}s_{r}}_{qh^{-1},kh^{-1},ph^{-1}}\\
= & T^{\left(H\right)s_{k}}_{k}\left(g^{-1},h;\sigma_{g},\sigma_{h}\right)+T^{\left(H\right)s_{p}}_{kg}\left(1,gh;1,\sigma_{h}\sigma_{g}\right)+T^{\left(H\right)s_{q}}_{k}\left(g^{-1}h^{-1},1;\sigma_{g}\sigma_{h},1\right)
\end{align*}
where
\[
T^{\left(H\right)s_{k}s_{p}s_{q}}_{kpq}=\text{Re}\left[s_{k}kQ\left(s_{p}p-s_{q}q\right)\chi^{s_{k}s_{p}s_{q}}_{kpq}\right]
\]
\[
T^{\left(H\right)s_{k}}_{k}\left(g^{-1},h,\sigma_{h},\sigma_{h}\right)=\text{Re}\left[k^{2}Q\left(\sigma_{p}g^{-1}-\sigma_{q}h\right)\chi^{s_{k}}_{k}\left(g^{-1},h;\sigma_{g}\sigma_{h}\right)\right]\;\text{.}
\]
Considering the equations of each of the other wave-numbers {[}e.g.
equations of the form $\partial_{t}E^{s_{p}}_{p}=T^{s_{p}s_{q}s_{k}}_{pqk}${]},
we can write:
\begin{align}
\Pi\left(k+\Delta k\right) & =\partial_{t}\int^{\infty}_{k+\Delta k}E\left(k'\right)dk'=\partial_{t}\sum_{k',s'>k}E^{s'}_{k'}=T^{s_{q}s_{k}s_{p}}_{qkp}-T^{s_{k}s_{p}s_{r}}_{k,\frac{k^{2}}{p},\frac{kq}{p}}\nonumber \\
 & =T^{s_{q}s_{k}s_{p}}_{qkp}-T^{s_{k}s_{p}s_{r}}_{k,kg,kgh}=T^{s_{q}s_{k}s_{p}}_{qkp}-T^{s_{k}s_{p}s_{r}}_{pg,kg,qg}\label{eq:flp}\\
 & =\text{Re}\bigg\{ s_{k}kQ\left[\left(1-\sigma_{g}g^{-1}\right)\chi^{s_{k}s_{p}s_{q}}_{kpq}-\sigma_{g}g\left(\sigma_{h}h-1\right)\chi^{s_{k}s_{p}s_{r}}_{pg,kq,qg}\right]\bigg\}\label{eq:Piptt}
\end{align}
here the integration is performed over $k'$, only over the modes
in the triad triplet, which can be interpreted for example as dirac
delta functions in $k$-space, so that the integral reduces to the
sum.

This is the the flux through a single triad triplet, and it considers
a single class of triads, so that in fact $\Pi=\Pi^{s_{k},\sigma_{g}\sigma_{h}}_{gh}\left(k+\Delta k\right)$
denoting the flux over the wave-number $k$ and helicity sign $s_{k}$
through the triad shape implied by $g$ and $h$ and the triad class
implied by by $\sigma_{g}$ and $\sigma_{h}$. 

Furthermore it is defined at $k+\Delta k$, and not at $k$, since
if we define the flux instead at $k-\Delta k$, it becomes:
\begin{align}
\Pi\left(k-\Delta k\right) & =-\partial_{t}\int^{k-\Delta k}_{0}E\left(k'\right)dk=T^{s_{k}s_{r}s_{q}}_{k,\frac{kp}{q},\frac{k^{2}}{q}}-T^{s_{p}s_{q}s_{k}}_{pqk}\nonumber \\
 & =T^{s_{k}s_{r}s_{q}}_{k,kh^{-1}g^{-1},kh^{-1}}-T^{s_{p}s_{q}s_{k}}_{pqk}=T^{s_{k}s_{r}s_{q}}_{qh^{-1},ph^{-1},kh^{-1}}-T^{s_{p}s_{q}s_{k}}_{pqk}\label{eq:flm}\\
 & =\text{Re}\bigg\{ ks_{k}Q\left[h^{-1}\sigma_{h}\left(1-\sigma_{g}g^{-1}\right)\chi^{s_{k}s_{r}s_{q}}_{qh^{-1},ph^{-1},kh^{-1}}-\left(\sigma_{h}h-1\right)\chi^{s_{k}s_{p}s_{q}}_{kpq}\right]\bigg\}\label{eq:Pimtt}
\end{align}
with
\begin{equation}
\partial_{t}\lim_{\Delta k\rightarrow0}\left(\frac{1}{2\Delta k}\int^{k+\Delta k}_{k-\Delta k}E\left(k\right)dk\right)=\lim_{\Delta k\rightarrow0}\left(\frac{\Pi\left(k-\Delta k\right)-\Pi\left(k+\Delta k\right)}{2\Delta k}\right)\;\text{,}\label{eq:lim}
\end{equation}
so that we get:
\begin{equation}
\partial_{t}E\left(k\right)=-\partial_{k}\Pi\left(k\right)\;\text{,}\label{eq:dap}
\end{equation}
where $\lim_{\Delta k\rightarrow0}\left(\frac{1}{\Delta k}E^{s_{k}}_{k}\right)\rightarrow E\left(k\right)$.

Note that a constant flux implies: $\Pi\left(k+\Delta k\right)=\Pi\left(k-\Delta k\right)$
such that:
\begin{equation}
\sigma_{h}h^{-1}\left(1-\sigma_{g}g^{-1}\right)\chi^{s_{k}s_{r}s_{q}}_{k,ph^{-1},qh^{-2}}+\sigma_{g}g\left(\sigma_{h}h-1\right)\chi^{s_{k}s_{p}s_{r}}_{k,pg^{2},qg}=\left(\sigma_{h}h-\sigma_{g}g^{-1}\right)\chi^{s_{k}s_{p}s_{q}}_{kpq}\;\text{,}\label{eq:threeeq}
\end{equation}
which can be satisfied for the following two scalings
\begin{equation}
\chi^{s_{k}s_{p}s_{q}}_{kpq}=\sigma_{g}g\chi^{s_{k}s_{p}s_{r}}_{pg,kg,qg}=\sigma_{h}h^{-1}\chi^{s_{k}s_{r}s_{q}}_{qh^{-1},ph^{-1},kh^{-1}}\label{eq:chisol1}
\end{equation}
\begin{equation}
\chi^{s_{k}s_{p}s_{q}}_{kpq}=g^{2}\chi^{s_{k}s_{p}s_{r}}_{pg,kg,qg}=h^{-2}\chi^{s_{k}s_{r}s_{q}}_{qh^{-1},ph^{-1},kh^{-1}}\label{eq:chisol2}
\end{equation}
The first solution implies $-1$ scaling for the $\chi_{kpq}$'s,
consistent with the Kolmogorov $-5/3$ power law solution. The flux
implied by this solution can be written as:
\begin{align}
\Pi\left(k-\Delta k\right)=\Pi\left(k+\Delta k\right) & =kQ\left[2-g^{-1}\left(\sigma_{g}+\sigma_{h}gh\right)\right]\text{Re}\left(s_{k}\chi^{s_{k}s_{p}s_{q}}_{kpq}\right)\;\text{,}\label{eq:pim}
\end{align}
whose direction depends on the sign of $\left[2-g^{-1}\left(\sigma_{g}+\sigma_{h}gh\right)\right]$
and of $s_{k}\cos\phi^{s_{k}s_{p}s_{q}}_{kpq}$. Note that since $g,h>1$,
as long as $\sigma_{h}=-1$ this term is necessarily positive. On
the other hand, if $\sigma_{g}$ and $\sigma_{h}$ are both positive
(i.e. homochiral interactions), the term is negative if $g^{-1}+h<2$
(e.g. $k=1.0$, $p=0.8$, $q=1.1$) and positive otherwise (e.g. $k=1.0$,
$p=0.8$, $q=1.5$), for class 3 (i.e. $\sigma_{h}=+1$, $\sigma_{g}=-1$),
the sign of this term is again always positive for any valid triangle.
Since the condition for it to become negative can be written as $h>2+g^{-1}$.
However it is impossible to satisfy this, since $h<2$ is necessary
in order for $k$, $p$ and $q$ to form a triad.

It is interesting to note that this is not consistent with the literature,
which states that the flux changes direction for sufficiently elongated
triads for the class 3\citep{waleffe:92,rathmann:17}. We think that
this behavior would be restored when we consider a full grid of triad
triplets, since in that case some of the energy that goes initially
to smaller scales, comes back to larger scales following the structure
of the grid, and the net flow can thus change direction because of
the elongation. 

Similarly the second solution implies a $-2$ scaling for the triple
correlations, consistent with the $-7/3$ scaling for $E\left(k\right)$,
with a flux defined as:
\begin{equation}
\Pi\left(k-\Delta k\right)=\Pi\left(k+\Delta k\right)=kQ\left[1-\sigma_{g}\sigma_{h}hg^{-1}\right]\text{Re}\left(s_{k}\chi^{s_{k}s_{p}s_{q}}_{kpq}\right)\;\text{,}\label{eq:pim2}
\end{equation}
whose direction is positive if $\sigma_{g}$ and $\sigma_{h}$ have
opposite signs and $s_{k}\cos\phi^{s_{k}s_{p}s_{q}}_{kpq}>0$. If
on the other hand $\sigma_{g}=\sigma_{h}$, the sign of the coefficient
changes sign if $\frac{qp}{k^{2}}>1$. 

At this point, we can make the connection to Kraichnan's formal similarity
analysis on his paper on two dimensional turbulence \citep{kraichnan:67}.
Kraichnan uses scaling factors, $v$ and $w$, which are closely related
to our $g$ and $h$ here, and integrating over both these geometric
factors, and over $k'$ (as we do in our Eqn. \ref{eq:flp}) for example
from $k$ up to infinity in order to compute the flux. The difference
in our formulation is that in some sense, because we do not really
integrate over $g$ and $h$, but organize the sums in such a way
that the cancellations due to the same shape/class triangle is already
explicitly considered, in the end we can write down a simplified expression
for the flux driven by a single shape/class, with cancellations from
the same shape/class already built-in. 

\section{Exact Solutions\label{sec:Exact-Solutions}}

\subsection{Power Law Solutions:\label{subsec:Exact-Power-Law}}

An interesting feature of the triad triplet is that if we fix the
triadic phases of the three triads that constitute it, well known
power law solutions for the spectral energy density would also satisfy
Eqn. (\ref{eq:tt}) directly. This can be seen by substituting a general
power law solution of the form:
\begin{equation}
u^{s_{k}}_{k}=s_{k}k^{-\alpha}e^{i\theta_{k}}=k^{-\alpha}e^{i\left(\theta_{k}+s_{k}\frac{\pi}{2}-\frac{\pi}{2}\right)}\rightarrow k^{-\alpha}e^{i\theta^{s_{k}}_{k}}\label{eq:sol1}
\end{equation}
in Eqn. \ref{eq:tthg}, and choosing $\phi_{k,p,q}=\phi_{\frac{k^{2}}{p},k,\frac{qk}{p}}=\phi_{\frac{k^{2}}{q},\frac{kp}{q},k}$
where $\phi_{kpq}=\theta_{k}+\theta_{p}+\theta_{q}$, so that each
of the three triads of the triplet have the same phase, which we can
call a ``chain synchronized'' solution (i.e. all the triads in a
triad chain, consisting of triplets are phase synchronized). Note
that the solution (\ref{eq:sol1}) requires the specific triadic phases
$\phi^{s_{k}s_{p}s_{q}}_{kpq}=\theta^{s_{k}}_{k}+\theta^{s_{p}}_{p}+\theta^{s_{q}}_{q}$
to be synchronized as follows:
\begin{equation}
\phi^{s_{k}s_{p}s_{q}}_{k,p,q}+s_{r}\pi/2=\phi^{s_{p}s_{k}s_{r}}_{\frac{k^{2}}{p},k,\frac{qk}{p}}+s_{q}\pi/2=\phi^{s_{q}s_{r}s_{k}}_{\frac{k^{2}}{q},\frac{kp}{q},k}+s_{p}\pi/2\;\text{.}\label{eq:sync1}
\end{equation}
Using (\ref{eq:sol1}) directly in (\ref{eq:tthg}) and dropping
the phases, with the above synchronization argument, we get:
\begin{align}
\partial_{t}u^{s_{k}}_{k}=s_{r}k^{1-2\alpha}Q\left(g,h\right)^{\sigma_{g}\sigma_{h}}\bigg[ & \left(1-\sigma_{g}g^{-1}\right)h^{2\alpha-1}g^{\alpha}+\left(\sigma_{g}g^{-1}-\sigma_{h}h\right)g^{\alpha}h^{-\alpha}\nonumber \\
 & +\left(\sigma_{h}h-1\right)g^{1-2\alpha}h^{-\alpha}\bigg]\;\text{,}\label{eq:vanish}
\end{align}
which vanishes when $2\alpha-1=-\alpha$ or $\alpha=1/3$.

Similarly a solution of the form:
\begin{equation}
u^{s_{i}}_{k}=k^{-\alpha}e^{i\theta^{s}_{k}}\label{eq:sol2}
\end{equation}
but this time with 
\begin{equation}
\phi^{s_{k}s_{p}s_{q}}_{k,p,q}=\phi^{s_{p}s_{k}s_{r}}_{\frac{k^{2}}{p},k,\frac{qk}{p}}=\phi^{s_{q}s_{r}s_{k}}_{\frac{k^{2}}{q},\frac{kp}{q},k}\label{eq:sync2}
\end{equation}
in other words with the phases being the same regardless of chirality,
when substituted into (\ref{eq:tthg}) gives:
\begin{align}
\partial_{t}u^{s_{k}}_{k}=s_{k}k^{1-2\alpha}Q\left(g,h\right)^{\sigma_{g}\sigma_{h}}\bigg[ & \left(\sigma_{h}-\sigma_{g}\sigma_{h}g^{-1}\right)h^{2\alpha-1}g^{\alpha}+\left(\sigma_{g}g^{-1}-\sigma_{h}h\right)g^{\alpha}h^{-\alpha}\nonumber \\
 & +\left(\sigma_{g}\sigma_{h}h-\sigma_{g}\right)g^{1-2\alpha}h^{-\alpha}\bigg]\label{eq:vanish2}
\end{align}
which vanishes for $2\alpha-1=1-\alpha$ or $\alpha=2/3$. 

From the definition of spectral energy density, we note that $E_{s}\left(k\right)=\left|u^{s}_{k}\right|^{2}k^{-1}$,
so that these two power law solutions (i.e. $\alpha=1/3$ and $\alpha=2/3$)
correspond to $E\left(k\right)\propto k^{-5/3}$ and $E\left(k\right)\propto k^{-7/3}$,
the energy and helicity cascade solutions respectively. It is interesting
to observe that both of these solutions satisfy the triad triplet
equation (that is the equation for $k$) exactly, regardless of the
triad class that is considered.

Note however that these solutions are actually valid on what can be
called a chain of triplets, in the sense that they assume that we
have the same equation (\ref{eq:tthg}) for all the wave-numbers that
appear in the triad triplet for $k$, so that the energy can flow
from one triplet to the next and then to the next. For example, our
equation for $u^{s_{p}}_{p}$ should be:
\begin{align*}
\partial_{t}u^{s_{p}}_{p}=\sigma_{g}g^{-1}s_{k}kQ\left(g,h\right)^{\sigma_{g}\sigma_{h}}\bigg[ & h^{-1}\sigma_{h}\left(1-\sigma_{g}g^{-1}\right)u^{s_{r}*}_{kg^{-1}h^{-1}}u^{s_{q}*}_{kh^{-1}g^{-2}}+\left(\sigma_{g}g^{-1}-\sigma_{h}h\right)u^{s_{k}*}_{kg^{-2}}u^{s_{r}*}_{khg^{-1}}\\
 & +g\sigma_{g}\left(\sigma_{h}h-1\right)u^{s_{q}*}_{kh}u^{s_{k}*}_{k}\bigg]\;\text{,}
\end{align*}
which can be obtained by transforming (\ref{eq:tthg}) via $k,s_{k}\rightarrow p,s_{p}$
instead of just 
\begin{equation}
\partial_{t}u^{s_{p}}_{p}=s_{k}kQ\left(g,h\right)^{\sigma_{g}\sigma_{h}}\left(\sigma_{h}h-1\right)u^{s_{q}*}_{kh}u^{s_{k}*}_{k}\label{eq:t2b}
\end{equation}
equivalent to Eqn (\ref{eq:t2}), so that when we substitute one of
the self-similar solutions discussed above, it satisfies the above
sets of equations for all wave-numbers, implying a network of connected
triad triplets, as will be discussed in section \ref{sec:network}.

However if one considers an actual ``isolated'' triad triplet, then
one has to satisfy boundary equations such as (\ref{eq:t2b}) with
given power law solutions that give $\partial_{t}u^{s_{k}}_{k}=0$.
This is possible for example if $u^{s_{q}}_{q},u^{s_{p}}_{p}\propto\left\{ U^{s_{q}}_{q}e^{\lambda^{s_{k}s_{p}s_{q}}_{kpq}t},U^{s_{p}}_{p}e^{\lambda^{s_{k}s_{p}s_{q}*}_{kpq}t}\right\} $,
with imaginary $\lambda^{s_{k}s_{p}s_{q}}_{kpq}$ so that $\lambda^{s_{k}s_{p}s_{q}}_{kpq}+\lambda^{s_{k}s_{p}s_{q}*}_{kpq}=0$.

\[
\left(\lambda^{s_{k}s_{p}s_{q}}_{kpq}\right)^{2}=\sigma_{h}k^{2}Q^{2}\left(1-\sigma_{g}g^{-1}\right)\left(h-\sigma_{h}\right)\left|u^{s_{k}}_{k}\right|^{2}\;\text{,}
\]
which gives an imaginary eigenvalue (hence an oscillating in time
solution), for $p$ and $q$ if $\sigma_{h}<0$, basically consistent
with the observation based on the instability assumption. If $\sigma_{h}>0$
on the other hand, $p$ and $q$ would grow in time, since the triad
is $k$-unstable.

For the smaller triad, we know that $k$ can not be a pump, which
means the eigenvalue:
\[
\left(\lambda^{s_{k}s_{q}s_{r}}_{k,qh^{-2},ph^{-1}}\right)^{2}=-h^{-2}k^{2}Q^{2}\left(h-\sigma_{h}\sigma_{g}g^{-1}\right)\left(h-\sigma_{h}\right)\left|u^{s_{k}}_{k}\right|^{2}
\]
is always imaginary since the right hand side is always negative.

Finally for the larger triad, we have
\[
\left(\lambda^{s_{k}s_{p}s_{r}}_{k,pg^{2},qg}\right)^{2}=-\sigma_{h}g^{2}k^{2}Q^{2}\left(h-\sigma_{g}\sigma_{h}g^{-1}\right)\left(1-\sigma_{g}g^{-1}\right)\left|u^{s_{k}}_{k}\right|^{2}
\]
which is imaginary if $\sigma_{h}>0$ and real if $\sigma_{h}<0$,
basically opposite of the middle triad.

Note that by adding damping terms to the unstable nodes. For example
for $\sigma_{h}<0$, considering 
\[
\partial_{t}u^{s_{p}}_{kg}=-\gamma_{d}u^{s_{p}}_{kg}-\sigma_{h}\sigma_{g}gs_{k}kQ\left(h-\sigma_{g}\sigma_{h}g^{-1}\right)u^{s_{k}*}_{k}u^{s_{r}*}_{khg}
\]
\[
\partial_{t}u^{s_{r}}_{kgh}=-\gamma_{d}u^{s_{r}}_{kgh}+\sigma_{g}gs_{k}kQ\left(1-\sigma_{g}g^{-1}\right)u^{s_{p}*}_{kg}u^{s_{k}*}_{k}
\]
with
\[
\gamma_{d}=gkQ\sqrt{\left(h-\sigma_{g}\sigma_{h}g^{-1}\right)\left(1-\sigma_{g}g^{-1}\right)}\left|u^{s_{k}}_{k}\right|
\]
one can make $u^{s_{p}}_{kg}$ and $u^{s_{r}}_{kgh}$ constant in
time. This is very similar to what one would normally call ``eddy
damping'' in this picture, chosen exactly to balance the energy that
is transferred to this pair. Normally in a connected chain of triplets,
this is to be balanced by the outgoing flux to the neighboring modes.

\subsection{Exact time dependent solutions on a triad triplet\label{subsec:Exact-time-dependent}}

\begin{figure}
\begin{centering}
\includegraphics[width=0.98\textwidth]{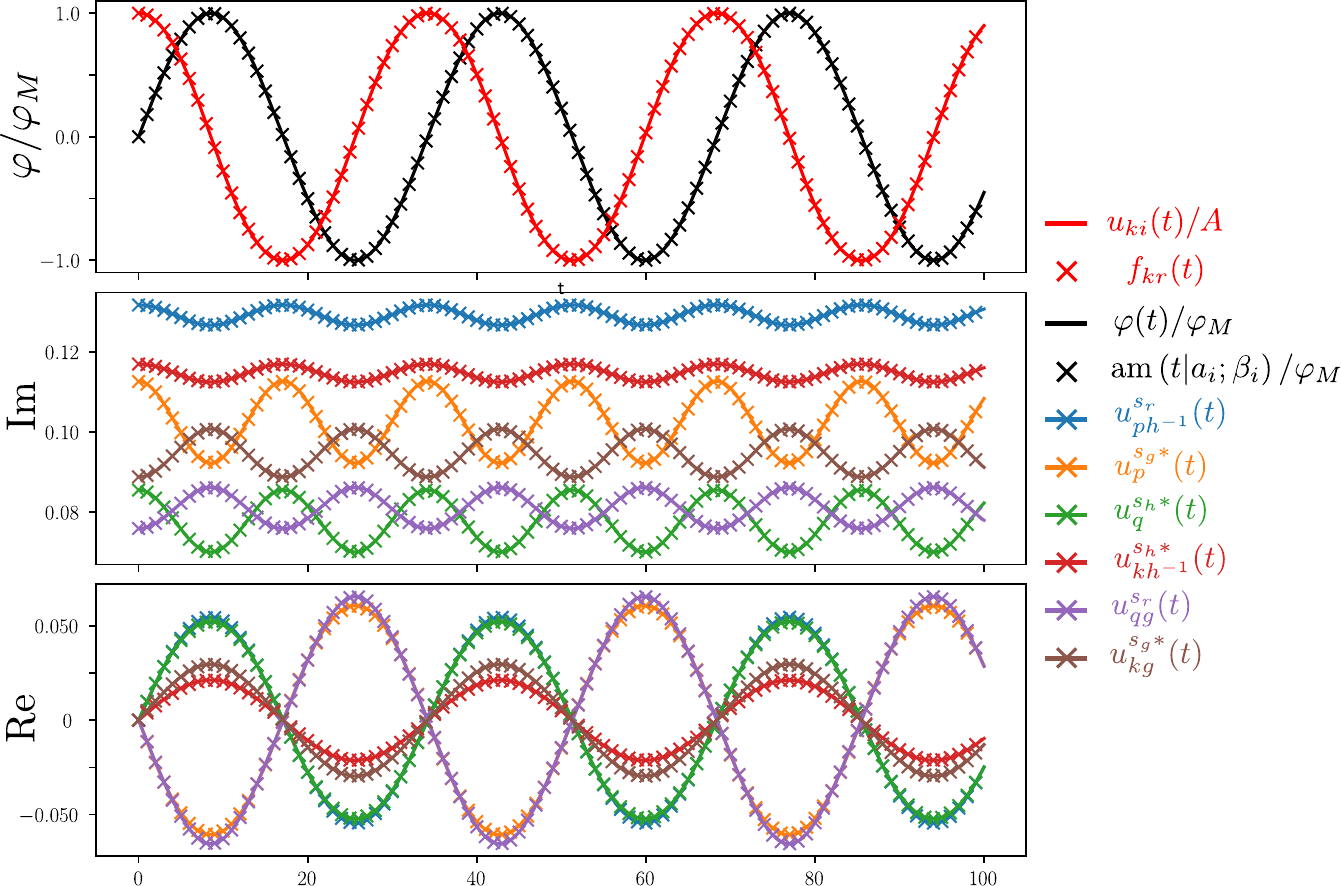}
\par\end{centering}
\caption{\label{fig:exact}Exact solution compared to the numerical solution
for $g=10/7$ and $h=8/5$, $\sigma_{h}=\sigma_{g}=-1$, and $A=0.1$.
Here $\varphi\left(t\right)$ is either obtained numerically using
$\varphi\left(t\right)=\beta^{-1}_{0}\arcsin\left[-\sigma_{h}u^{s_{q}}_{kh^{-1}r}\left(t\right)\sqrt{\lambda_{h}}/\left(Ag^{1/3}h^{1/3}\sqrt{\lambda_{h}+\sigma_{h}\lambda_{g}}\right)\right]$,
or analytically from $\varphi\left(t\right)=\text{am}^{\sigma_{h}}\left(t|a_{i};\beta_{i}\right)$
, by inverting Eqn. \ref{eq:incomp}. The functions $f_{i}\left(t\right)$
and $f_{kr}\left(t\right)$are then computed using Eqn. \ref{eq:sols},
and Eqn. \ref{eq:fkr} respectively, which are then subsituted into
\ref{eq:usols} to obtain the solutions. Here, we plot the conjugates
of the solutions with negative imaginary parts in order to show them
together in a single plot. Top plot shows the phase $\varphi$ normalized
to its maximum value $\varphi_{M},$and the solution $u_{ki}\left(t\right)/A$,
and the bottom two plots show the imaginary and real parts of the
solutions for the rest of the wavenumbers. The numerical solution
is computed by solving the equations of the triad triplet (e.g. Eqn.
\ref{eq:tteqs}) numerically, with the same initial conditions as
the exact ones. For reference, with these parameters we have, $\beta_{0}=0.05$,
$\beta_{1}=0.11$, $\beta_{2}=0.0926$, and $a_{0}=1.445$, $a_{1}=-0.0521$,
$a_{2}=3.159$, which give $\varphi_{M}=5.544$ and $\tau=8.548$.}
\end{figure}

Consider the equations of an isolated triad triplet as written explicitly
in the appendix \ref{sec:appA}. Defining the coefficients 
\[
\alpha\equiv s_{k}kQ^{s_{k}s_{p}s_{q}}_{kpq}
\]
\begin{align*}
\lambda_{g} & \equiv\left|1-\sigma_{g}g^{-1}\right|=1-\sigma_{g}g^{-1}\\
\lambda_{h} & \equiv\left|\sigma_{h}h-1\right|=h-\sigma_{h}
\end{align*}
so that we have
\[
\lambda_{g}\sigma_{h}+\lambda_{h}=h-\sigma_{h}\sigma_{g}g^{-1}=\left|\sigma_{h}h-\sigma_{g}g^{-1}\right|
\]
and that we can write the triad triplet equations as:
\begin{align}
\partial_{t}u^{s_{k}}_{k}= & \alpha\sigma_{h}\bigg[h^{-1}\lambda_{g}u^{s_{q}*}_{kh^{-1}}u^{s_{r}*}_{ph^{-1}}-\left(\lambda_{g}\sigma_{h}+\lambda_{h}\right)u^{s_{p}*}_{p}u^{s_{q}*}_{q}\nonumber \\
 & \qquad+g\sigma_{g}\lambda_{h}u^{s_{r}*}_{qg}u^{s_{p}*}_{kg}\bigg]\nonumber \\
\partial_{t}u^{s_{p}}_{p}= & \alpha\sigma_{h}\lambda_{h}u^{s_{q}*}_{q}u^{s_{k}*}_{k}\nonumber \\
\partial_{t}u^{s_{q}}_{q}= & \alpha\lambda_{g}u^{s_{k}*}_{k}u^{s_{p}*}_{p}\nonumber \\
\partial_{t}u^{s_{r}}_{ph^{-1}}= & \alpha h^{-1}\lambda_{h}u^{s_{k}*}_{k}u^{s_{q}*}_{kh^{-1}}\nonumber \\
\partial_{t}u^{s_{q}}_{kh^{-1}}= & -\alpha h^{-1}\left(\lambda_{g}\sigma_{h}+\lambda_{h}\right)u^{s_{r}*}_{ph^{-1}}u^{s_{k}*}_{k}\nonumber \\
\partial_{t}u^{s_{r}}_{qg}= & \alpha\sigma_{g}g\lambda_{g}u^{s_{p}*}_{kg}u^{s_{k}*}_{k}\nonumber \\
\partial_{t}u^{s_{p}}_{kg}= & -\alpha\sigma_{g}\sigma_{h}g\left(\lambda_{g}\sigma_{h}+\lambda_{h}\right)u^{s_{k}*}_{k}u^{s_{r}*}_{qg}\;\text{.}\label{eq:tteqs}
\end{align}
In order to find an exact, time dependent solution, let us start by
proposing a solution of the form:
\begin{align}
u^{s_{k}}_{k} & =Af_{kr}e^{i\pi/2}\nonumber \\
u^{s_{r}}_{ph^{-1}} & =A\left(f_{0r}-i\sqrt{\frac{\lambda_{h}}{\lambda_{h}+\sigma_{h}\lambda_{g}}}g^{-1/3}f_{0i}\right)g^{1/3}h^{1/3}e^{i\sigma_{g}\sigma_{h}\pi/2}\nonumber \\
u^{s_{q}}_{kh^{-1}} & =A\left(f_{0r}+i\sqrt{\frac{\lambda_{h}+\sigma_{h}\lambda_{g}}{\lambda_{h}}}g^{1/3}f_{0i}\right)h^{1/3}e^{i\sigma_{h}\pi/2}\nonumber \\
u^{s_{p}}_{p} & =A\left(f_{1r}+i\sigma_{h}\sqrt{\frac{\lambda_{h}}{\lambda_{g}}}g^{-1/3}h^{-1/3}f_{1i}\right)g^{1/3}e^{i\sigma_{g}\pi/2}\nonumber \\
u^{s_{q}}_{q} & =A\left(f_{1r}+i\sqrt{\frac{\lambda_{g}}{\lambda_{h}}}g^{1/3}h^{1/3}f_{1i}\right)h^{-1/3}e^{i\sigma_{h}\pi/2}\nonumber \\
u^{s_{p}}_{kg} & =A\left(f_{2r}+i\sqrt{\frac{\left(\lambda_{h}+\sigma_{h}\lambda_{g}\right)}{\lambda_{g}}}h^{-1/3}f_{2i}\right)g^{-1/3}e^{i\sigma_{g}\pi/2}\nonumber \\
u^{s_{r}}_{qg} & =A\left(f_{2r}-i\sigma_{h}\sqrt{\frac{\lambda_{g}}{\left(\lambda_{h}+\sigma_{h}\lambda_{g}\right)}}h^{1/3}f_{2i}\right)g^{-1/3}h^{-1/3}e^{i\sigma_{h}\sigma_{g}\pi/2}\label{eq:usols}
\end{align}
which gives:
\begin{align*}
\partial_{t}f_{0r} & =\alpha A\sigma_{g}h^{-1}\sqrt{\left(\lambda_{h}+\sigma_{h}\lambda_{g}\right)\lambda_{h}}f_{0i}f_{kr}\\
\partial_{t}f_{0i} & =-\alpha A\sigma_{g}h^{-1}\sqrt{\lambda_{h}\left(\lambda_{h}+\sigma_{h}\lambda_{g}\right)}f_{0r}f_{kr}\\
\partial_{t}f_{1r} & =\alpha A\sigma_{g}\sqrt{\lambda_{h}\lambda_{g}}f_{1i}f_{kr}\\
\partial_{t}f_{1i} & =\alpha A\sigma_{h}\sigma_{g}\sqrt{\lambda_{h}\lambda_{g}}f_{1r}f_{kr}\\
\partial_{t}f_{2r} & =\alpha A\sigma_{g}\sigma_{h}g\sqrt{\lambda_{g}\left(\lambda_{h}+\sigma_{h}\lambda_{g}\right)}f_{2i}f_{kr}\\
\partial_{t}f_{2i} & =-\alpha A\sigma_{g}g\sqrt{\lambda_{g}\left(\lambda_{h}+\sigma_{h}\lambda_{g}\right)}f_{2r}f_{kr}
\end{align*}
for the real and imaginary parts of the functions $f_{n}=f_{n}\left(t\right),$with
$n=0,1,2$ and $f_{k}=f_{kr}\left(t\right)$ as the normalized time
dependent functions, that make up our solution. Taking the initial
conditions as $f_{nr}=1$, while $f_{ni}=0$, we find the relations
$f^{2}_{0r}+f^{2}_{0i}=1$, $f^{2}_{1r}-\sigma_{h}f^{2}_{1i}=1$ and
$f^{2}_{2r}+\sigma_{h}f^{2}_{2i}=1$. Note that while this seems restrictive,
the numerical solution of the original triad triplet equations with
these initial conditions actually follow the exact solutions that
we find here (see Figure \ref{fig:exact}). We can then consider a
slightly more general class of initial conditions by moving the time
offset.

Using these relations we can link the imaginary parts of $f_{i}$
functions to $f_{kr}$ as:
\begin{equation}
\frac{\partial_{t}f_{0i}}{\beta_{0}\sqrt{1-f^{2}_{0i}}}=\frac{\partial_{t}f_{1i}}{\beta_{1}\sqrt{1+\sigma_{h}f^{2}_{1r}}}=\frac{\partial_{t}f_{2i}}{\beta_{2}\sqrt{1-\sigma_{h}f^{2}_{2i}}}=f_{kr}\label{eq:dts}
\end{equation}
where we have defined
\[
\beta_{0}\equiv-\alpha A\sigma_{g}h^{-1}\sqrt{\lambda_{h}\left(\lambda_{h}+\sigma_{h}\lambda_{g}\right)}
\]
\[
\beta_{1}=\alpha A\sigma_{h}\sigma_{g}\sqrt{\lambda_{h}\lambda_{g}}
\]
\[
\beta_{2}=-\alpha A\sigma_{g}g\sqrt{\lambda_{g}\left(\lambda_{h}+\sigma_{h}\lambda_{g}\right)}
\]
We can ``solve'' Eqn. \ref{eq:dts} by choosing:
\begin{align}
f_{0i} & =\sin\left(\beta_{0}\varphi\right)\nonumber \\
f_{1i} & =-\sigma_{h}\sqrt{-\sigma_{h}}\sin\left(\sqrt{-\sigma_{h}}\beta_{1}\varphi\right)\nonumber \\
f_{2i} & =\sigma_{h}\sqrt{\sigma_{h}}\sin\left(\sqrt{\sigma_{h}}\beta_{2}\varphi\right)\label{eq:sols}
\end{align}
note that for $\sigma_{h}=1$, $f_{1i}=-i\sin\left(i\beta_{1}\varphi\right)=\sinh\left(\beta_{1}\varphi\right)$,
while $f_{2i}=\sin\left(\beta_{2}\varphi\right)$, while for $\sigma_{h}=-1$,
we have $f_{1i}=\sin\left(\beta_{1}\varphi\right)$ but $f_{2i}=\sinh\left(\beta_{2}\varphi\right)$,
underlining the role of $\sigma_{h}$ once again defining the triad
stability. Here, the triad that is $k$-unstable, has a hyperbolic
sine. For Eqn. \ref{eq:sols} to solve Eqns. \ref{eq:dts}, means
that we must also have:

\begin{equation}
f_{kr}=\partial_{t}\varphi\;\text{.}\label{eq:solk}
\end{equation}
To solve it, we need $f_{kr}$ in terms of $\varphi$, and we can
obtain such a relation by using energy and helicity conservation to
link $f_{k}$ to the other three functions $f_{n}$. One would normally
think this would give two independent relations, but due to the degeneracy
of choosing to represent the other two wavenumbers of each triad using
the same function, they actually only give one relation, which can
be written as:
\begin{align*}
f^{2}_{kr} & =1-a_{0}f^{2}_{0i}-a_{1}f^{2}_{1i}-a_{2}f^{2}_{2i}
\end{align*}
with $a_{0}\equiv\sigma_{h}\left[\frac{1}{\lambda_{h}}g^{2/3}-\frac{1}{\lambda_{h}+\sigma_{h}\lambda_{g}}\right]\lambda_{g}h^{2/3}$,
$a_{1}\equiv\left[\frac{h^{-2/3}}{\lambda_{g}}+\sigma_{h}\frac{g^{2/3}}{\lambda_{h}}\right]\left(\lambda_{h}+\sigma_{h}\lambda_{g}\right)$
and $a_{2}\equiv\left[\frac{h^{-2/3}}{\lambda_{g}}-\frac{\sigma_{h}}{\left(\lambda_{h}+\sigma_{h}\lambda_{g}\right)}\right]\lambda_{h}g^{-2/3}$.

When substituted into Eqn. \ref{eq:solk}, this gives some kind of
generalization of the elliptic integral, that we can write as:
\begin{equation}
t\left(\varphi\right)=\int^{\varphi}_{0}\frac{d\varphi'}{\sqrt{1-a_{0}\sin^{2}\left(\beta_{0}\varphi\right)+a_{1}\sigma_{h}\sin^{2}\left(\sqrt{-\sigma_{h}}\beta_{1}\varphi\right)-a_{2}\sigma_{h}\sin^{2}\left(\sqrt{\sigma_{h}}\beta_{2}\varphi\right)}}\;\text{.}\label{eq:incomp}
\end{equation}
Notice that the maximum and the minimum values of $f^{2}_{kr}$ are
$1$ and $0$ which happen at $t=0$ and $t=\tau$ respectively, with
$\tau$ being the quarter period, defined as
\begin{equation}
\tau=\int^{\varphi_{M}}_{0}\frac{d\varphi'}{\sqrt{1-a_{0}\sin^{2}\left(\beta_{0}\varphi'\right)+a_{1}\sigma_{h}\sin^{2}\left(\sqrt{-\sigma_{h}}\beta_{1}\varphi'\right)-a_{2}\sigma_{h}\sin^{2}\left(\sqrt{\sigma_{h}}\beta_{2}\varphi'\right)}}\;\text{,}\label{eq:per}
\end{equation}
where $\varphi_{M}$ can be defined by the relation:
\begin{equation}
1-a_{0}\sin^{2}\beta_{0}\varphi_{M}+a_{1}\sigma_{h}\sin^{2}\left(\sqrt{-\sigma_{h}}\beta_{1}\varphi_{M}\right)-a_{2}\sigma_{h}\sin^{2}\left(\sqrt{\sigma_{h}}\beta_{2}\varphi_{M}\right)=0\;\text{.}\label{eq:phim}
\end{equation}

These functions, which appear to be generalizations of Jacobi elliptic
functions to 3+3 parameters (i.e. the coefficients $a,b,c$ and the
scaling factors $\beta_{i}$), involving two $\sin^{2}$ and one $\sinh^{2}$
terms in the denominator, can be implemented numerically by solving
Eqn. \ref{eq:phim} to obtain $\varphi_{M}$, then using the generalized
complete integral of Eqn. \ref{eq:per} to compute the quarter period
$\tau$ (equivalent to $\pi/2$ for trigonometric functions), and
finally using the generalized incomplete integral in Eqn. \ref{eq:incomp}
to compute $t$ as a function of $\varphi$, which can then be inverted
numerically to find $\varphi=\varphi\left(t\right)$. This resulting
function, which is the properly periodized inverse of Eqn. \ref{eq:incomp}
is equivalent to the Jacobi amplitude $\varphi\left(t\right)=\text{am}^{\sigma_{h}}\left(t|a_{i};\beta_{i}\right)$,
which can easily be extended to $t>\tau$ using its periodicity. Then
the solutions are given by Eqns. \ref{eq:sols} and \ref{eq:solk}.
Note that we can also use 
\begin{equation}
f_{kr}\left(t\right)=\left(-1\right)^{\left(\left\lfloor t/\tau\right\rfloor +\left\lfloor t/2\tau\right\rfloor \right)}\sqrt{1-a_{0}\sin^{2}\left(\beta_{0}\varphi'\right)+a_{1}\sigma_{h}\sin^{2}\left(\sqrt{-\sigma_{h}}\beta_{1}\varphi'\right)-a_{2}\sigma_{h}\sin^{2}\left(\sqrt{\sigma_{h}}\beta_{2}\varphi'\right)}\label{eq:fkr}
\end{equation}
explicitly, to obtain $u^{s_{k}}_{k}\left(t\right)$ instead of Eqn.
\ref{eq:solk}.

The properly periodized function $\varphi\left(t\right)=\text{am}^{\sigma_{h}}\left(t|a_{i};\beta_{i}\right)$,
is implemented in python, and can be used to write the exact periodic
solutions of the triad triplet equations, whose comparison to numerical
solutions of the \ref{eq:tteqs} with the same initial conditions
are shown in Figure \ref{fig:exact} for $g=10/7$ and $h=8/5$, $\sigma_{h}=\sigma_{g}=-1$,
with $k=1$,$s_{k}=1$ and $A=0.1$, so that $\alpha=Q=0.524$, $\beta_{0}=0.05$,
$\beta_{1}=0.11$, $\beta_{2}=0.0926$, and $a_{0}=1.445$, $a_{1}=-0.0521$,
$a_{2}=3.159$, which in turn give $\varphi_{M}=5.544$ and $\tau=8.548$.
The function works as expected for all the values of $g$, $h$, $\sigma_{g}$
and $\sigma_{h}$ that we tested.

Notice that, these solutions are periodic in time, generating no energy
or helicity transfer. Or rather, the energy that is transferred to
smaller scales bounces right back and then forth because it has nowhere
else to go. In this sense, the energy in these solutions can be considered
as being ``trapped'' in the finite $k$-space of the triplet, as
opposed to the constant flux solution where the energy would be ``passing''
through the triplet that is connected to other triplets as a chain
or a network. Recall that the initial condition has a $\left|u_{k}\left(0\right)\right|\propto k^{-1/3}$
scaling with the initial phases chosen to satisfy the condition for
the corresponding exact solution on a chain (i.e. Eqn. \ref{eq:sync1}).
We can write a similar solution for initial conditions with a $k^{-2/3}$
scaling instead of $k^{-1/3}$ with the phases chosen according to
Eqn. \ref{eq:sync2}. However when we compute time averages, because
the oscillating generalized snoidal functions do not have the same
averages, one does not get the same power law scaling for the mean
values as the initial scaling.

Note finally that, these are indeed a \emph{new class of exact nonlinear
solutions} of the Navier-Stokes equations, which can be written in
real space as:
\[
\mathbf{u}\left(\mathbf{x},t\right)=\sum_{\mathbf{k}\in\mathbf{k}_{i}}\hat{\mathbf{h}}^{s'}_{\mathbf{k}'}u^{s_{k'}}_{\mathbf{k}'}\left(t\right)e^{i\mathbf{k}'\cdot\mathbf{r}}
\]
where the $\hat{\mathbf{h}}^{s}_{\mathbf{k}}$ are the eigenmodes
of the curl operator in Fourier space defined via $i\mathbf{k}\times\hat{\mathbf{h}}^{s}_{\mathbf{k}}=sk\hat{\mathbf{h}}^{s}_{k}$.
Writing these solutions explicitly in cartesian coordinates in three
dimensional real space, they should also be useful for benchmarking
nonlinear simulations \citep{ethier:94}. When plotted in real space
these solutions evoke tilted periodic patterns that move about each
other periodically with a nonlinear frequency determined by the shape
and class of the triad as well as the overall amplitude as given by
Eqn. \ref{eq:per}. The picture of energy trapping in $k$-space suggests
that the structures become slightly smaller and slightly larger during
different phases of these oscillations, as thay move about.

\section{Network topography of triadic interactions:\label{sec:network}}

It is tempting to imagine that putting triad triplets ``end to end''
we could construct a linear chain of triads over which the energy
flows. Such a picture is not completely wrong, especially in the sense
that considering triad triplets at one scale, then the next scale
etc., we can trace how the energy or helicity can be transferred in
Fourier space, suggesting that the triad triplet can be considered
as a local diagnostic tool, which allows us to focus on the role of
a single shape/class of triad in the overall cascade.

However when we look at how the triad triplet construct actually expands
in $k$-space as we add more triads to each of its end nodes the resulting
structure appears to be more complex than a linear chain. The obvious
way to expand the triad triplet would be to write evolution equations
for all the nodes that appear on the right hand side of for example
Eqn. (\ref{eq:tthg}). One can write the equation for a node, say
$p,s_{p}$ basically by applying the affine transformation to the
equation for $k$, $s_{k}$ given in Eqn. (\ref{eq:tthg}). 

Note that an isolated triad triplet can be described by a network
consisting of the $7$ nodes (dropping helical modes and directions
for simplicity) labeled by:
\begin{equation}
k_{i}=k\left\{ gh,g,h,1,g^{-1},h^{-1},g^{-1}h^{-1}\right\} \label{eq:nodes}
\end{equation}
\begin{figure}
\begin{centering}
\includegraphics[width=0.99\textwidth]{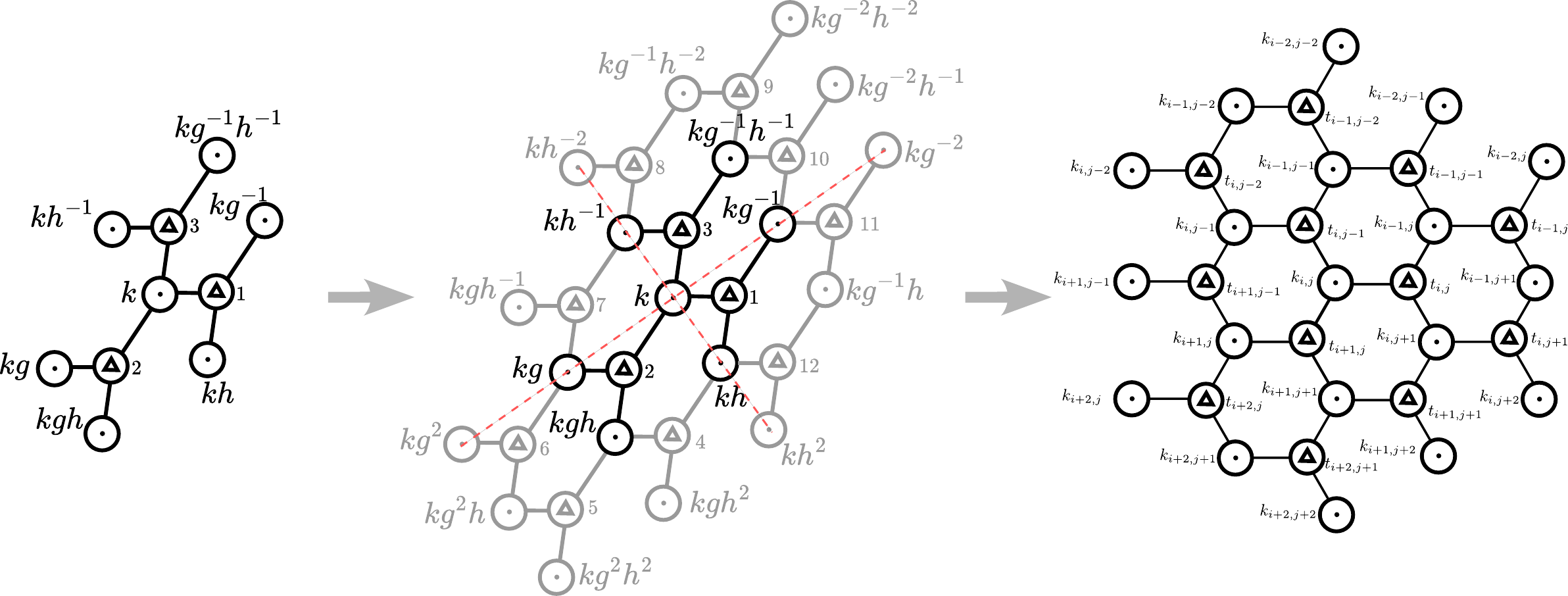}
\par\end{centering}
\caption{\label{fig:nw}Bipartite network representation of the triad triplet,
where each node represented by a point is connected to a triad represented
by a small triangle, and each triad connects to only three nodes.
The graph shows how the triad triplet connects to the rest of the
wave-number space network of triads of the same shape. Here each node
is taken to represent all the helical modes (these can also be separated)
as well as different orientations. As we add more triplets, instead
of forming a linear chain, the network forms a hexagonal mesh because
each node is connected to three triads and each triad is connected
to three nodes. Note how the nodes and the triangles are organized
in diagonal lines indicated by the red (if in color) dashed line.
The resulting grid can be written as $k_{ij}=kg^{i}h^{j}$ , and the
triads can be reorganized so that we can label them as $t_{ij}=k\left[g^{i}h^{j},g^{i-1}h^{j},g^{i}h^{j+1}\right]$.}
\end{figure}
and the $3$ triads:
\begin{equation}
\left\{ t_{1},t_{2},t_{3}\right\} =k\left\{ \left[1,g^{-1},h\right],\left[g,1,gh\right],\left[h^{-1},g^{-1}h^{-1},1\right]\right\} \;\text{.}\label{eq:triads}
\end{equation}
This forms a hypergraph, which can be represented as a bipartite network
using the wave-number space network formulation\citep{gurcan:20,gurcan:23},
shown as the left hand plot of figure \ref{fig:nw}. In order to grow
this isolated triplet into a full network, we take each node in this
list, and consider the triad triplet that has that node as its middle
element (or equivalently the three triads that take this node first
as the smallest, than as the middle and then as the largest leg) with
the same triad shape (so that we have the same $h$ and $g$). This
would give us a triad ``dodecuplet'' that can be seen in figure
\ref{fig:nw}, which consists of $19$ nodes and $12$ triads. The
procedure can be repeated until we cover all the Fourier space.

Note that the resulting mesh can actually be written as $k_{ij}=kg^{i}h^{j}$
, and the triads can be reorganized so that we can label the triad
involving the node $k_{ij}$ as the middle wave number and $t_{ij}$
as:
\[
t_{ij}=k\left[g^{i}h^{j},g^{i-1}h^{j},g^{i}h^{j+1}\right]\;\text{,}
\]
and if we somehow have $g^{n}=h^{m}$ for some $n$ and $m$ integer,
the grid becomes ``periodic'' with $k_{i+n,j}=k_{i,j+m}$, resulting
in a spiral chain in wave-number space. In the case of $g=h$, we
get a shell model like topology, or a butterfly instead of a triplet
as the basic element, using the terminology developed in the wave
turbulence community\citep{kartashova:08,lvov:09}.

Notice also that since each triad has exactly three nodes, and each
node on a network of triad triplets has exactly three triads connected
to it, we have a complementarity between triads and nodes. This suggests
a hexagonal grid structure for the triad network, as can be seen in
figure \ref{fig:nw}, and the triad triplet can be seen as a compact
stencil on such a grid. Furthermore, the complementarity allows one
to switch to a representation where the triads become the primary
elements and nodes represent three body interactions among those.
This makes sense especially when considering the three correlations
$\chi_{kpq}$, or the triadic phases, which decide the fate of the
energy flux through a given triad.

\begin{figure}
\begin{centering}
\includegraphics[width=0.5\textwidth]{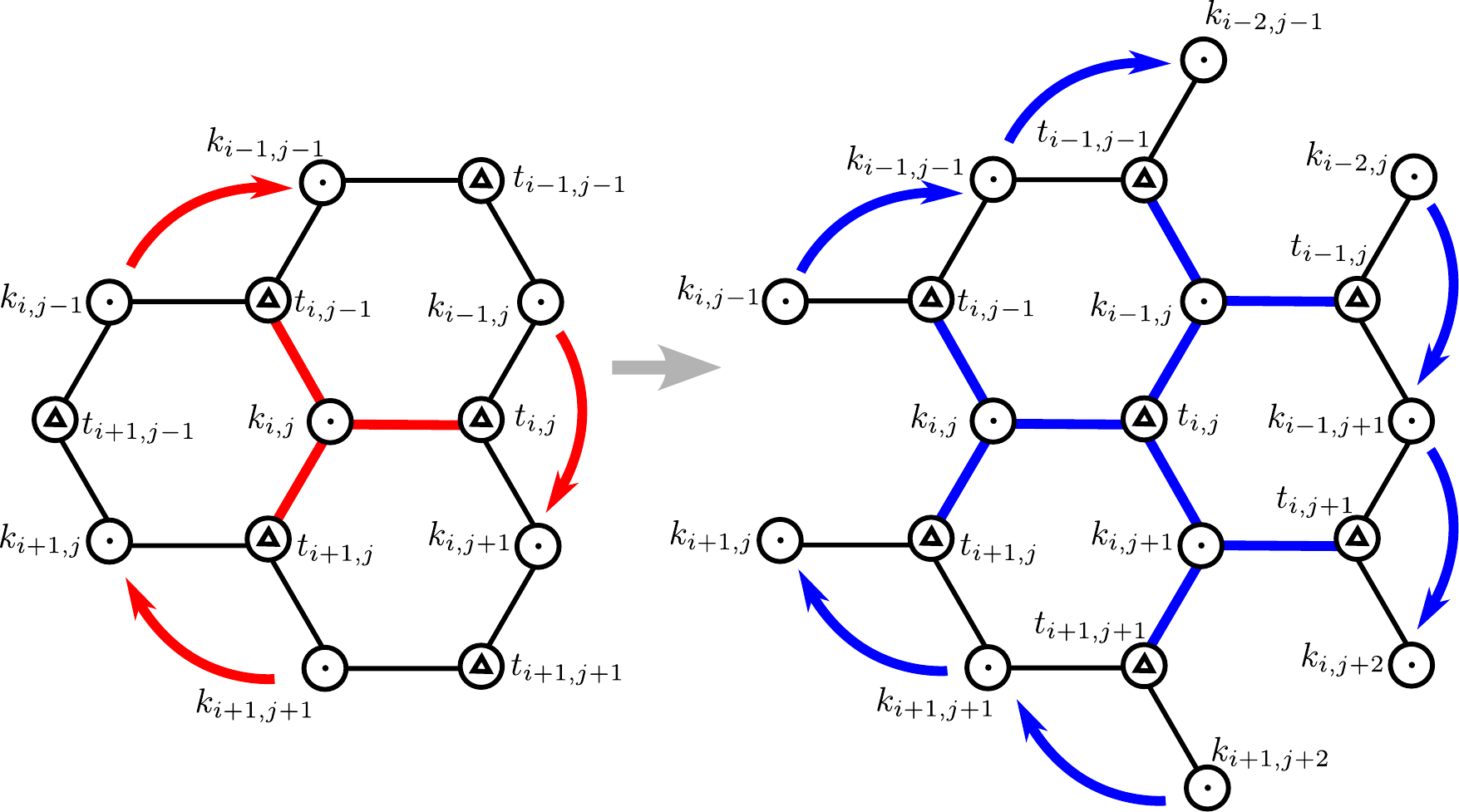}
\par\end{centering}
\caption{\label{fig:hex}The diagramatic description of Eqns. (\ref{eq:eneq})
on the left and Eqn. (\ref{eq:pheq}) on the right. The thick solid
(red if in color) lines without arrows indicate the triads to which
the node $k_{ij}$ is connected, the directions of the arrows (which
is always clockwise in this construction) indicate the order of the
coefficients as $\left(s'k'-s''k''\right)$ where $s'$, $k'$ is
the node from which the arrow originates and $s'',k''$ is the node
towards which the arrow points. The thick solid (blue if in color)
lines without arrows on the right indicate the connections of the
triad $t$ to other triads $t'$ (via the nodes that connect them)
and the arrows indicate the coefficients $\left(s'k'-s''k''\right)$
the same way, with the rest of the coefficient being a property either
of the triad $t'$ or of the node $k$ that connects the two triads.}
\end{figure}

We can rewrite Eqn. (\ref{eq:tthg}) with an arbitrary mesh labeling
as:
\begin{align}
\partial_{t}u^{s_{ij}}_{i,j}=Q\left(g,h\right)^{\sigma_{g}\sigma_{h}}\bigg[ & \left(s_{i,j-1}k_{ij-1}-s_{i-1,j-1}k_{i-1,j-1}\right)u^{s_{ij-1}*}_{i,j-1}u^{s_{i-1,j-1}*}_{i-1,j-1}\nonumber \\
 & +\left(s_{i-1,j}k_{i-1,j}-s_{i,j+1}k_{i,j+1}\right)u^{s_{i-1,j}*}_{i-1,j}u^{s_{i,j+1}*}_{i,j+1}\nonumber \\
 & +\left(s_{i+1,j+1}k_{i+1,j+1}-s_{i+1,j}k_{i+1,j}\right)u^{s_{i+1,j+1}*}_{i+1,j+1}u^{s_{i+1,j}*}_{i+1,j}\bigg]\;\text{,}\label{eq:tt_gen}
\end{align}
or diagramatically as:
\begin{equation}
\partial_{t}u^{s}_{k}=s\sum_{t}Q^{\sigma_{g}\sigma_{h}}_{t}\left(\sigma_{g}k'-\sigma_{h}k''\right)u^{s'*}_{k'}u^{s''*}_{k''}\label{eq:udiag}
\end{equation}
where the sum is over the triads $t$ that are connected to the node
$k$ (with the helicity sign $s$) and $s'$,$k'$ and s'',$k''$
are the other two nodes of each triad as shown in Figure \ref{fig:hex},
with the given class $\sigma_{g},\sigma_{h}$. Different classes can
be included in the sum over triads basically using the same notation
either explicitly or by considering each triad class as a different
``triad node'' in the network formulation. One can also include
different shapes and orientations in such a diagrammatic representation.
However our goal here is not to recover general network formulation
of turbulence but rather focus on a network consisting of a single
kind of triad, so that the equations for the other elements of the
grid can be obtained by scaling and rotating Eqn (\ref{eq:tt_gen}),
which corresponds to increasing or decreasing either $i$ or $j$
and together with the parity transformation $s_{j}\rightarrow s_{\ell}$
to get the parity that is needed so that we stay on a mesh made up
of affine transformations of one triad with a given shape and class.
As such, this mesh provides a ``mono-fractal'', self-affine basis
of interacting triads, along which the turbulent cascade can happen
depending on the phases of its elements. This provides the basis for
a simple percolation on a self-affine hypergraph perspective of the
turbulent cascade that we discuss in section \ref{subsec:perc} below.

In contrast in real turbulence there are many different kinds of triads,
and one may argue that it can be represented by a network of interacting
triad triplets. Two such networks would interact with one another
only at their common nodes, which might be relatively sparse. However
on a discrete $k$-space grid corresponding to a bounded system, all
these different chains would be forced to coexist on the same discrete
grid and therefore all of them will have common nodes at large scales.
But also at small scales rounding may result in $g^{n}=h^{m}$ giving
rise to periodicity. This enforces a more or less tree-like graph
structure at large scales that branches into asymptotically self-affine
networks made up of different triads at smaller scales, which may
have certain reconnecting branches again at smaller scales.

The energy equations can be written on wave-number nodes as:
\begin{equation}
\partial_{t}E^{s}_{k}=\sum_{t}Q_{t}\left(s'k'-s''k''\right)\left|\chi^{ss's''}_{t}\right|\cos\phi^{ss's''}_{t}\label{eq:eneq}
\end{equation}
where $\chi^{ss's''}_{t}=u^{s}_{k}u^{s'}_{k'}u^{s''}_{k''}$ and the
sum is over the triads that are connected to the node $k$. Note that
since the grid is constructed by affine transformations, which follows
a chain made up of single shape and class of triads, whether we write
the interactions explicitly in terms of the ratios $g$, $h$, $\sigma_{g}$,
$\sigma_{h}$, or in this case in terms of $\left(s',k';s'',k''\right)$'s
as in Eqn. (\ref{eq:eneq}), the result is the same.

The equation for the $\chi^{ss's''}_{t}$, which are triple correlations
that are naturally associated with the triads can be written by multiplying
the equation (\ref{eq:udiag}) with $u^{s'}_{k'}$ and $u^{s''}_{k''}$
and considering the permutations of $k,s\rightarrow k',s'\rightarrow k'',s''$,
which gives:
\begin{align}
\partial_{t}\chi_{t}= & Q_{t}\sum_{k\in\left\{ t,t'\right\} ,\left\{ k',k''\right\} \in t'}\frac{\left(s'k'-s''k''\right)}{E^{s}_{k}}\chi^{*}_{t'}\chi_{t}\label{eq:chiteq}
\end{align}
where the label $t$ represents $\chi_{t}=\chi^{s,s',s''}_{kpq}=\chi^{s,\sigma_{g}s,\sigma_{h}s}_{k,kg^{-1},kh}$,
and the sum over $t'$ is over the $6$ triads that are connected
to each of the three nodes of $t$ such that $k,s$ is the node considered
in the sum, and $k'$, $s'$ and $k''$, $s''$ are the other nodes
that appear in $t'$ in the order of permutation $k,s\rightarrow k',s'\rightarrow k'',s''$
as shown in figure \ref{fig:hex}. Note that the sum also includes
$t$, and $t$ appears three times in the sum because it shares three
nodes with itself. If we write the self-contribution of $t$ explicitly,
we get:
\begin{align}
\partial_{t}\chi_{t}= & Q_{t}\bigg\{\left(s'k'-s''k''\right)E^{s'}_{k'}E^{s'}_{k''}+\left(s''k''-sk\right)E^{s''}_{k''}E^{s}_{k}\nonumber \\
 & +\left(sk-s'k'\right)E^{s}_{k}E^{s'}_{k'}+\sum_{t\neq t'}\frac{\left(s'k'-s''k''\right)}{E^{s}_{k}}\chi^{*}_{t'}\chi_{t}\bigg\}\label{eq:chiteq2}
\end{align}
The equation for the triad phases can be obtained from (\ref{eq:chiteq})
by using $\chi_{t}=\left|\chi_{t}\right|e^{i\phi_{t}}$, multiplying
by $e^{-i\phi_{t}}$ and taking the imaginary part as:
\begin{equation}
\partial_{t}\phi_{t}=Q_{t}\sum_{t'}\left|\chi_{t'}\right|\frac{\left(s'k'-s''k''\right)}{E^{s}_{k}}\sin\phi_{t'}\label{eq:pheq}
\end{equation}
with the sum containing the self interaction terms as in (\ref{eq:chiteq2}).
Note also that taking the imaginary part of (\ref{eq:chiteq2}) directly,
we get the alternative form of the phase equations that take the form
\begin{align}
\partial_{t}\left(\left|\chi_{t}\right|\sin\phi_{t}\right)= & Q_{t}\bigg\{\sum_{t\neq t'}\frac{\left(s'k'-s''k''\right)}{E^{s}_{k}}\chi^{*}_{t'}\chi_{t}\sin\left(\phi_{t}-\phi_{t'}\right)\bigg\}\label{eq:pheq2}
\end{align}
where the self\_contributions from $t$ drop because they are real.
It is easy to see that a fixed point of (\ref{eq:chiteq2}) is $\phi_{t}=\phi_{t'}$
with $\sin\phi_{t}=m_{t}/\left|\chi_{t}\right|$, where $m_{t}$ is
a constant related to the initial phases of each triad.

Finally, it is worth noting that the general form of these equations
suggests a percolation structure on a three body network (i.e. a hypergraph),
where the energies, and helicities, associated with the wave-number
nodes can be transferred from one node to another if the triads that
connect them have the right phases. See section \ref{subsec:perc}
for a more detailed discussion.

\subsection{Connection to percolation\label{subsec:perc}}

As discussed in Section \ref{sec:single_triad}, the four triad classes
correspond to $\sigma_{g},\sigma_{h}=\left(+,+\right),\left(+,-\right),\left(-,+\right),\left(-,-\right)$,
that we label from $1$ to $4$, and the instability assumption determines
the fate of the energy that is injected in one of the legs of each
of these triads such that, for an isolated triad, the energy goes
from the wavenumber that has the overall sign of its nonlinear interaction
coefficient that is different from the other two, towards the others,
which for three dimensional Navier-Stokes turbulence depends solely
on the sign of $\sigma_{h}$. Considering this wave-number as the
``pump'', we argue that if there are multiple triads that are connected
to a given wave-number the energy is transferred through the triad
where the wave-number node in question acts as the pump. 
\begin{figure}
\begin{centering}
\includegraphics[width=0.5\textwidth]{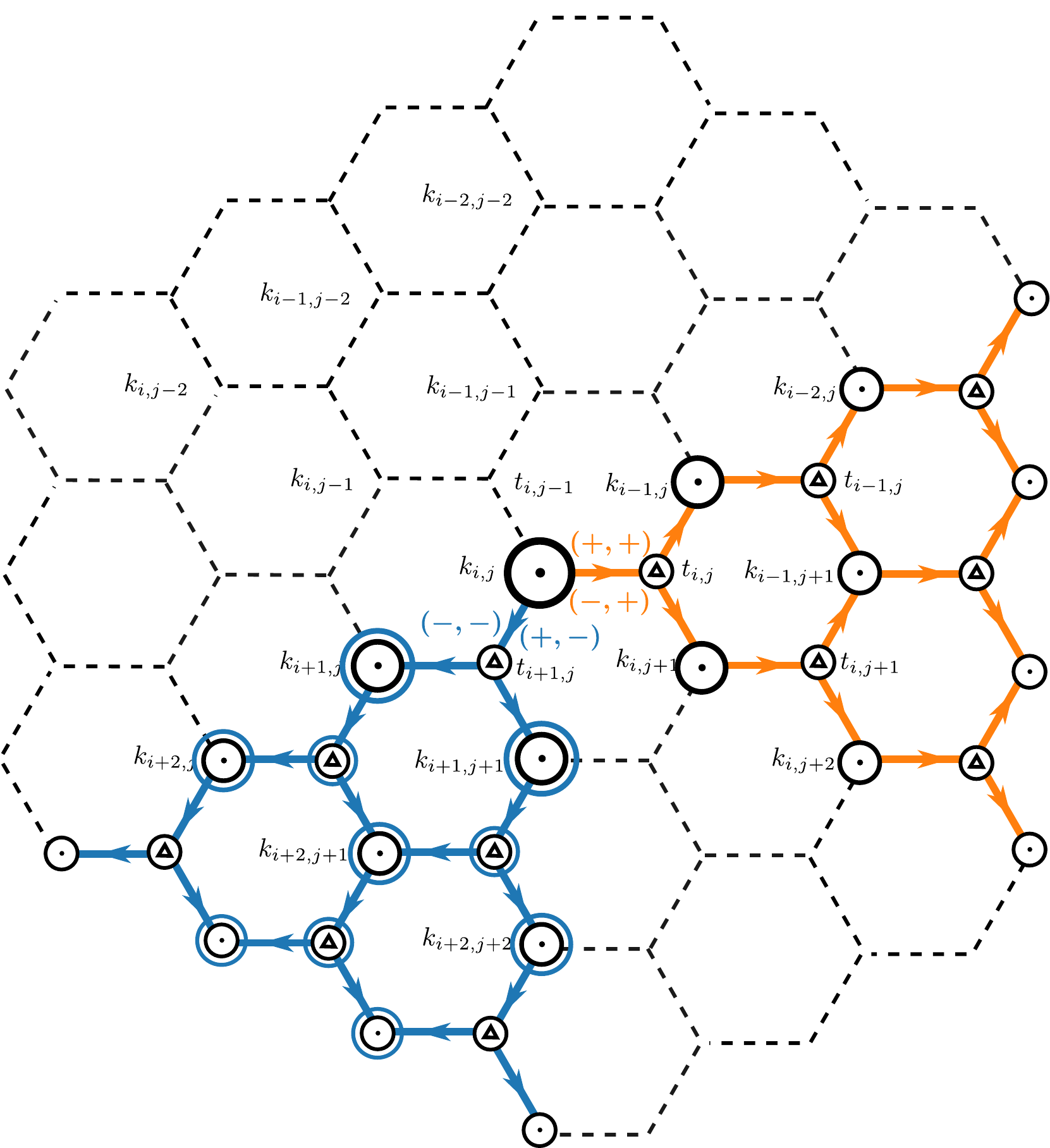}
\par\end{centering}
\caption{\label{fig:hextrees} The tree structures for different classes, where
the energy goes in a different direction depending on the class. When
the two larger wavenumbers have the same chirality, the energy goes
from the middle leg to the other two as in 2D, while if the two larger
wavenumbers have the opposite chirality, the energy goes from the
smallest wavenumber towards the larger two wavenumbers. In either
case, these connections form tree-like structures in a hexagonal grid,
where a triad triplet centered at $k_{i+2,j+1}$, forming a compact
stencil on this hexagonal grid, is highlighted by circling its nodes
and triads for the forward cascade tree.}
\end{figure}

Note that a simple cartoon of energy transfer is that the energy goes
from the pump to the other two wave-numbers in each triad of the web
discussed in \ref{sec:network}. Starting from a given wave-number,
say $k_{ij}$ the energy will flow in the direction depending on the
class as seen in figure \ref{fig:hextrees}, basically going from
the initial wavenumber $k_{ij}$ towards $t_{ij}$ in the case of
classes 1 and $3$ and towards $t_{i+1,j}$ in the case of classes
2 and 4. 

This allows the energy injected in $k_{ij}$ to fill up the corresponding
triangular domain of the hexagonal grid. Considering the phase relations
(or resonant interactions in the case of wave-turbulence) as generating
either open or closed links on such a domain, we can map the problem
of the turbulent cascade in $k$-space through a single triad shape/class
to that of directed percolation on a hexagonal grid. In the same spirit,
the more general case of turbulent cascade on an arbitrary set of
triads can be formulated as that of percolation on a hypergraph, where
the phase relation determine if the flux can go through a given hyperlink.
One can construct a simple discrete system where at each time step,
the energy at a given pump mode $k$ would be reduced by $\Delta E_{k}$
and is sent to the two other modes $p$ and $q$ proportional to the
$\Delta E_{p}/\Delta E_{k}$ and $\Delta E_{q}/\Delta E_{k}$ as computed
in section \ref{sec:single_triad}. Unfortunately, the details of
such a description is beyond the scope of the current paper.

Finally, using Figure \ref{fig:hextrees} as a map, we can also consider
the evolution of a finite tree of activated triads. For example considering
$\sigma_{h}<0$ so that we have a forward cascade, we can consider
the three triads consisting of $t_{i+1,j}$, $t_{i+2,j}$ and $t_{i+2,j+1}$
with their six wave-numbers, and attempt to solve it both numerically
and analytically as we have done in this paper for the three triads
$t_{i,j}$, $t_{i+1,j}$ and $t_{i,j-1}$ and seven wave-numbers of
the triad triplet.

\section{Conclusion\label{sec:Conclusion}}

Starting from the dynamics of a single triad in helical decomposition,
and noting that the geometric factor of the triad depends only on
the ratios of the largest (i.e. $q$) and smallest (i.e. $p$) wave-numbers
to the middle wave-number $k$, which defines its shape, and the relative
chiralities of those to that of the middle wave-number (i.e. $s_{q}s_{k}$
and $s_{p}s_{k}$) , which defines its class, it was observed that
considering three consecutive triads of the same shape and class,
where the reference wave-number $k$ appears as the smallest, the
middle and the largest wave-numbers respectively, gives a particularly
interesting object called the triad triplet. Since the triad itself
can be defined by two affine transformations of a reference wave-number,
in the sense that the first transformation takes $k$, $s_{k}$ and
produces $p$, $s_{p}$ and the second transformation takes $k$,
$s_{k}$ and produces $q$, s$_{q}$, it is possible to construct
the evolution equation of the triad triplet by applying the inverses
of these affine transformations on the equations for $p$, $s_{p}$
and $q,s_{q}$ in order to consider the contribution of the same triad
shape/class to the evolution of $k$, where $k$ is the smallest,
middle and largest wave-numbers of each of the triads that constitute
the triplet.

One result of the paper is that, well known solutions, such as the
energy cascade solution $E\left(k\right)\propto k^{-5/3}$ and the
helicity cascade solution $E\left(k\right)\propto k^{-7/3}$, satisfy
the triad triplet equations on an infinite chain of triad triplets,
regardless of the shape or the class of the triad. The direction of
the cascade can be determined by the triad instability assumption
together with a simple comparison of the nonlinear interaction coefficients,
which determine the effective inertia of each mode. It is noted that
the energy injected in a given node, would fill up the corresponding
triangular domain of the hexagonal grid, depending on the sign of
$\sigma_{h}$. It is argued that, considering the cosines of the phases
as approaching $+1$, $0$ and $-1$, the turbulent cascade can be
linked to directed percolation in a topologically hexagonal mesh of
wave-number and triad nodes. It is further noted that the more general
problem of the energy cascade in an arbitrary network of nodes in
wave-number domain can be mapped, again by replacing the phases by
ones and zeros to percolation on a hypergraph.

Beyond these observations, the key result is the identification of
a class of exact solutions of the Navier-Stokes equations, by solving
the equations for an isolated triad triplet, using a generalization
of the Jacobi elliptic functions. In real space, these solutions appear
in the form of tilted periodic patterns that move around and become
larger and smaller periodically with a nonlinear frequency given by
Eqn. \ref{eq:per}. In Fourier space, they can be thought of as trapped
solutions where the energy in the system is trapped in the triads
constituting the triplet.

\appendix

\section{Solution of single triad equations in terms of elliptic functions\label{sec:appA}}

Generally speaking the solutions of (\ref{eq:t1}-\ref{eq:t3}) can
be written, for instance as: 
\begin{align}
u^{s_{k}}_{k}\left(t\right) & =\sqrt{\frac{\left(\frac{q}{k}-s_{p}s_{q}\frac{p}{k}\right)}{\left(\frac{q}{k}-s_{q}s_{k}\right)}}u^{s_{p}}_{p}\left(0\right)\text{sn}\left(\omega t,\kappa\right)e^{i\left(\phi_{p}-\phi_{q}\right)}\label{eq:uks}\\
u^{s_{p}}_{p}\left(t\right) & =u^{s_{p}}_{p}\left(0\right)\text{cn}\left(\omega t,\kappa\right)e^{i\left(\phi_{q}-\phi_{k}\right)}\nonumber \\
u^{s_{q}}_{q}\left(t\right) & =u^{s_{q}}_{q}\left(0\right)\text{dn}\left(\omega t,\kappa\right)e^{i\left(\phi_{k}-\phi_{p}\right)}\nonumber 
\end{align}
where
\begin{equation}
\omega=k\left|Q^{s_{k}s_{p}s_{q}}_{kpq}\right|\sqrt{\left(\frac{q}{k}-s_{q}s_{k}\right)\left(\frac{q}{k}-s_{p}s_{q}\frac{p}{k}\right)}u^{s_{q}}_{q}\left(0\right)\label{eq:om}
\end{equation}
and 
\begin{equation}
m=s_{q}s_{k}\frac{\left(1-s_{k}s_{p}\frac{p}{k}\right)\left|u^{s_{p}}_{p}\left(0\right)\right|^{2}}{\left(\frac{q}{k}-s_{q}s_{k}\right)\left|u^{s_{q}}_{q}\left(0\right)\right|^{2}}\;\text{.}\label{eq:m}
\end{equation}
Note that for $\sigma_{h}\equiv s_{q}s_{k}<0$ this becomes negative,
which necessitates the use of analytical continuation of the Jacobi
elliptic functions for negative $m$. Similarly if $\left(1-s_{k}s_{p}\frac{p}{k}\right)\left|u^{s_{p}}_{p}\left(0\right)\right|^{2}>\left(\frac{q}{k}-s_{q}s_{k}\right)\left|u^{s_{q}}_{q}\left(0\right)\right|^{2}$,
then $m$ may become larger than $1$, which can be handled by the
Jacobi's real transformation. One may also note that a general solution
can be formulated in terms of the Weirstrass elliptic function, which
is actually connected to this general (analytically continued) form
of the Jacobi elliptic functions.

\section{full triad triplet equations:\label{sec:appB}}

The full triad triplet equations, for an isolated triplet can be defined
using the coefficients, explicitly in absolute values in order to
make their signs explicit as:

\begin{align*}
\partial_{t}u^{s_{k}}_{k}= & s_{k}kQ^{s_{k}s_{p}s_{q}}_{kpq}\bigg[h^{-1}\sigma_{h}\left|1-\sigma_{g}g^{-1}\right|u^{s_{q}*}_{kh^{-1}}u^{s_{r}*}_{kg^{-1}h^{-1}}-\sigma_{h}\left|\sigma_{h}h-\sigma_{g}g^{-1}\right|u^{s_{p}*}_{kg^{-1}}u^{s_{q}*}_{kh}\\
 & +g\sigma_{g}\sigma_{h}\left|\sigma_{h}h-1\right|u^{s_{r}*}_{kgh}u^{s_{p}*}_{kg}\bigg]\\
\partial_{t}u^{s_{p}}_{kg^{-1}}= & s_{k}kQ^{s_{k}s_{p}s_{q}}_{kpq}\sigma_{h}\left|\sigma_{h}h-1\right|u^{s_{q}*}_{kh}u^{s_{k}*}_{k}\\
\partial_{t}u^{s_{q}}_{kh}= & s_{k}kQ^{s_{k}s_{p}s_{q}}_{kpq}\left|1-\sigma_{g}g^{-1}\right|u^{s_{k}*}_{k}u^{s_{p}*}_{kg^{-1}}\\
\partial_{t}u^{s_{r}}_{kg^{-1}h^{-1}}= & s_{k}kQ^{s_{k}s_{p}s_{q}}_{kpq}\left|1-\sigma_{h}h^{-1}\right|u^{s_{k}*}_{k}u^{s_{q}*}_{kh^{-1}}\\
\partial_{t}u^{s_{q}}_{kh^{-1}}= & -s_{k}kQ^{s_{k}s_{p}s_{q}}_{kpq}\left|1-\sigma_{g}\sigma_{h}g^{-1}h^{-1}\right|u^{s_{r}*}_{kg^{-1}h^{-1}}u^{s_{k}*}_{k}\\
\partial_{t}u^{s_{r}}_{kgh}= & s_{k}kQ^{s_{k}s_{p}s_{q}}_{kpq}\sigma_{g}\left|\sigma_{g}g-1\right|u^{s_{p}*}_{kg}u^{s_{k}*}_{k}\\
\partial_{t}u^{s_{p}}_{kg}= & -s_{k}kQ^{s_{k}s_{p}s_{q}}_{kpq}\sigma_{g}\sigma_{h}\left|\sigma_{g}\sigma_{h}gh-1\right|u^{s_{k}*}_{k}u^{s_{r}*}_{kgh}
\end{align*}
where the first equation corresponds to the central node and the consecutive
pairs of equations correspond to the other two legs of the first (middle),
second (smaller) and third (larger) triads respectively. Since the
geometric factor $Q$ is always the same, and $s_{k}$ multiplies
all the equations, it is easy to see that if $\sigma_{h}>0$, $k$
looses energy to the primary triad, while it is $kh^{-1}$ and $kg$
which loose their energies in the second and third triads. For $\sigma_{h}<0$,
$k$ looses energy to the third triad, while it is the $kg^{-1}$
and $kg^{-1}h^{-1}$ which loose their energies in the first and second
triads.

\section{Triad Triplet Basis\label{sec:appC}}

Let us consider the Navier Stokes equation written in the following
form
\begin{align}
\partial_{t}u^{s_{k}}\left(\mathbf{k}\right) & +\sum_{s_{p},s_{q}}\iint\limits_{p<k<q}Q^{s_{k}s_{p}s_{q}}_{kpq}\left(s_{p}p-s_{q}q\right)u^{s_{p}*}\left(\mathbf{p}\right)u^{s_{q}*}\left(\mathbf{q}\right)\delta^{3}\left(\mathbf{k}+\mathbf{p}+\mathbf{q}\right)d^{3}\mathbf{p}d^{3}\mathbf{q}\nonumber \\
 & +\sum_{s_{p'},s_{q'}}\iint\limits_{k<q'<p'}Q^{s_{k'}s_{p'}s_{q'}}_{k'p'q'}\left(s_{p'}p'-s_{q'}q'\right)u^{s_{p'}*}\left(\mathbf{p}'\right)u^{s_{q'}*}\left(\mathbf{q}'\right)\delta^{3}\left(\mathbf{k}+\mathbf{p}'+\mathbf{q}'\right)d^{3}\mathbf{p}'d^{3}\mathbf{q}'\nonumber \\
 & +\sum_{s_{p''},s_{q''}}\iint\limits_{q''<p''<k}Q^{s_{k}s_{p''}s_{q''}}_{kp''q''}\left(s_{p''}p''-s_{q''}q''\right)u^{s_{p''}*}\left(\mathbf{p}''\right)u^{s_{q''}*}\left(\mathbf{q}''\right)\delta^{3}\left(\mathbf{k}+\mathbf{p}''+\mathbf{q}''\right)d^{3}\mathbf{p}''d^{3}\mathbf{q}''\nonumber \\
 & =-\nu k^{2}u^{s_{k}}\left(\mathbf{k}\right)\label{eq:nsgen}
\end{align}
where we seperate the integral explicitly into three regions, and
label the dummy integration variables differently so that we can choose
them explicitly in a way that will give the same geometric factor
$Q^{s_{k}s_{p}s_{q}}_{kpq}$.

As discussed in Section \ref{subsec:nsintt}, assuming $p<k<q$, we
can parametrize a triad using any two of the four variables $g,h,\alpha$
and $\beta$, plus its orientation $\phi$, and its class defined
by $\sigma_{g}$ and $\sigma_{h}$. This allows us to write the interaction
relative to the wave-number, helicity pair defined as $\left(\mathbf{k},s_{k}\right)$.

In order to parametrize a triad using $g$ and $\alpha$, we can first
set the orientation $\phi=0$, and use complex number notation for
two dimensional vectors, where the real part represents $\hat{\mathbf{x}}$
and imaginary part represents $\mathbf{\hat{y}}$. In this notation,
a wave-vector $\mathbf{k}$ in $\mathbf{\hat{y}}$ direction, which
has a sign of helicity $s_{k}$ would be represented by $\left(\mathbf{k},s_{k}\right)=\left(ik,s_{k}\right)$.
We can write $\left(\mathbf{p},s_{p}\right)=\left(ikz^{-1}_{g},s_{k}\sigma_{p}\right)$
and $\left(\mathbf{q},s_{q}\right)=\left(ikz_{h},s_{k}\sigma_{q}\right)$,
where $z_{g}\equiv ge^{i\alpha}$ and $z_{h}\equiv he^{i\beta}$,
which gives a triad with $p<k<q$ assuming $g$ and $h$ are both
greater than $1$.

Now in order to parametrize the triads with $k<q'<p'$, in a way that
it gives the same geometric factor, we define $\left(\mathbf{q}',s_{q'}\right)\equiv\left(ikz_{g},s_{p}\right)$
and $\left(\mathbf{p}',s_{p'}\right)\equiv\left(ikz_{g}z_{h},s_{r}\right)$.
This gives 
\begin{equation}
\hat{\mathbf{z}}\times\mathbf{p}'\cdot\mathbf{q}'=\text{Im}\left(\mathbf{p}'^{*}\mathbf{q}'\right)=\text{Im}\left(k^{2}g^{2}he^{-i\beta}\right)=-k^{2}g^{2}h\sin\beta=k^{2}gh\sin\left(\alpha+\beta\right)\label{eq:zppqp}
\end{equation}
where we used Eqn. \ref{eq:gh_ab}, and therefore
\begin{align}
Q^{s_{k}s_{p'}s_{q'}}_{kp'q'} & =\frac{1}{4}s_{k}s_{p'}s_{q'}\frac{\text{Im}\left(\mathbf{p}'^{*}\mathbf{q}'\right)}{p'q'}\left(s_{k}+s_{p'}\frac{p'}{k}+s_{q'}\frac{q'}{k}\right)\nonumber \\
 & =\frac{1}{4}s_{k}s_{r}s_{p}\frac{gh\sin\left(\alpha+\beta\right)}{g^{2}h}\left(s_{k}+s_{r}gh+s_{p}g\right)\nonumber \\
 & =\frac{1}{4}s_{k}s_{p}s_{q}\sin\left(\alpha+\beta\right)\left(s_{p}g^{-1}+s_{q}h+s_{k}\right)=Q^{s_{k}s_{p}s_{q}}_{kpq}\label{eq:qp}
\end{align}
looking at the rest of the interaction coefficient, we have:
\begin{equation}
\left(s_{p'}p'-s_{q'}q'\right)=kg\left(s_{r}h-s_{p}\right)\label{eq:nl2p}
\end{equation}
Similarly in order to parametrize the triads with $q''<p''<k$, we
choose $\left(\mathbf{p}'',s_{p''}\right)=\left(ikz^{-1}_{h},s_{q}\right)$
and $\left(\mathbf{q}'',s_{q''}\right)=\left(ikz^{-1}_{h}z^{-1}_{g},s_{r}\right)$,
which gives
\begin{equation}
\hat{\mathbf{z}}\times\mathbf{p}''\cdot\mathbf{q}''=\text{Im}\left(\mathbf{p}''^{*}\mathbf{q}''\right)=\text{Im}\left(k^{2}h^{-2}g^{-1}e^{-i\alpha}\right)=-k^{2}h^{-2}g^{-1}\sin\alpha=k^{2}g^{-1}h^{-1}\sin\left(\alpha+\beta\right)\label{eq:zppp}
\end{equation}
again using Eqn. \ref{eq:gh_ab}, and writing the geometric factor
explicitly, we get:
\begin{align}
Q^{s_{k}s_{p''}s_{q''}}_{kp''q''} & =\frac{1}{4}s_{k}s_{p''}s_{q''}\frac{\text{Im}\left(\mathbf{p}''^{*}\mathbf{q}''\right)}{p''q''}\left(s_{k}+s_{p''}\frac{p''}{k}+s_{q''}\frac{q''}{k}\right)\nonumber \\
 & =\frac{1}{4}s_{k}s_{q}s_{r}\frac{g^{-1}h^{-1}\sin\left(\alpha+\beta\right)}{g^{-1}h^{-2}}\left(s_{k}+s_{q}h^{-1}+s_{r}g^{-1}h^{-1}\right)\nonumber \\
 & =\frac{1}{4}s_{p}s_{k}s_{q}\sin\left(\alpha+\beta\right)\left(s_{q}h+s_{k}+s_{p}g^{-1}\right)=Q^{s_{k}s_{p}s_{q}}_{kpq}\label{eq:Qpp}
\end{align}
together with
\begin{equation}
\left(s_{p''}p''-s_{q''}q''\right)=kh^{-1}\left(s_{q}-s_{r}g^{-1}\right)\label{eq:nl2pp}
\end{equation}
Writing the components of $\mathbf{p}$ explicitly
\[
\mathbf{p}=kg^{-1}\sin\alpha\cos\phi\hat{\mathbf{x}}+kg^{-1}\cos\alpha\hat{\mathbf{y}}+kg^{-1}\sin\alpha\sin\phi\hat{\mathbf{z}}
\]
and $\mathbf{q}$ using $\mathbf{q}=-\mathbf{k}-\mathbf{p}$, after
computing the integral over $\mathbf{q}$.We can write the Jacobian
of the transformation as:
\begin{align*}
J & =\left|\frac{\partial\mathbf{p}}{\partial g}\times\frac{\partial\mathbf{p}}{\partial\alpha}\cdot\frac{\partial\mathbf{p}}{\partial\phi}\right|=k^{3}g^{-4}\sin\alpha
\end{align*}
and compute the volume element as:
\[
d^{3}\mathbf{p}=k^{3}g^{-4}\sin\alpha dgd\alpha d\phi=-k^{3}g^{-3}hdgdhd\phi\;\text{,}
\]
where we used Eqn (\ref{eq:alp1}) to write $\sin\alpha d\alpha=-hgdh$.

Similarly for the triads with $k<q'<p'$:
\[
\mathbf{q}'=kg\left(-\sin\alpha\cos\phi\hat{\mathbf{x}}+\cos\alpha\hat{\mathbf{y}}-\sin\alpha\sin\phi\hat{\mathbf{z}}\right)
\]
with $\mathbf{p}'=-\mathbf{k}-\mathbf{q}'$
\begin{align*}
J & =\frac{\partial\mathbf{q}'}{\partial g}\times\frac{\partial\mathbf{q}'}{\partial\alpha}\cdot\frac{\partial\mathbf{q}'}{\partial\phi}=k^{3}g^{2}\left(-\sin\alpha\cos\phi\hat{\mathbf{x}}+\cos\alpha\hat{\mathbf{y}}-\sin\alpha\sin\phi\hat{\mathbf{z}}\right)\times\left(-\cos\alpha\cos\phi\hat{\mathbf{x}}-\sin\alpha\hat{\mathbf{y}}-\cos\alpha\sin\phi\hat{\mathbf{z}}\right)\cdot\left(\sin\alpha\sin\phi\hat{\mathbf{x}}-\sin\alpha\cos\phi\hat{\mathbf{z}}\right)\\
 & =-k^{3}g^{2}\left(\sin^{3}\alpha\cos^{2}\phi+\cos^{2}\alpha\cos^{2}\phi+\sin^{3}\alpha\sin^{2}\phi+\cos^{2}\alpha\sin\alpha\sin^{2}\phi\right)=-k^{3}g^{2}\sin\alpha
\end{align*}
\[
J=\frac{\partial\mathbf{q}'}{\partial g}\times\frac{\partial\mathbf{q}'}{\partial\alpha}\cdot\frac{\partial\mathbf{q}'}{\partial\phi}=-k^{3}g^{2}\sin\alpha
\]
so that 
\[
d^{3}\mathbf{q}'=-k^{3}g^{2}\sin\alpha dgd\alpha d\phi=k^{3}g^{3}hdgdhd\phi
\]
and finally for the triads with $q''<p''<k$
\[
\mathbf{p}''=kh^{-1}\left(\sin\beta\cos\phi\hat{\mathbf{x}}+\cos\beta\hat{\mathbf{y}}+\sin\beta\sin\phi\hat{\mathbf{z}}\right)
\]
\[
J=\frac{\partial\mathbf{p}''}{\partial h}\times\frac{\partial\mathbf{p}''}{\partial\beta}\cdot\frac{\partial\mathbf{p}''}{\partial\phi}=k^{3}h^{-4}\sin\beta\;\text{,}
\]
which gives: 
\[
d^{3}\mathbf{p}''=k^{3}h^{-4}\sin\beta dhd\beta d\phi=k^{3}h^{-5}g^{-3}dgdhd\phi
\]
using Eqn (\ref{eq:bet1}), to write $\sin\beta d\beta=g^{-3}h^{-1}dg$. 

These choices allow us to write the integral as 
\begin{align*}
\partial_{t}u^{s_{k}}\left(\mathbf{k}\right) & +s_{k}k^{4}\int^{\infty}_{1}\int^{\infty}_{1}\int^{2\pi}_{0}\sum_{\sigma_{p},\sigma_{q}}Q\left(g,h\right)^{\sigma_{g}\sigma_{h}}\bigg[\\
 & \left(\sigma_{g}g^{-1}-\sigma_{h}h\right)u^{s_{p}*}\left(kg^{-1},-\alpha,\phi\right)u^{s_{q}*}\left(kh,\beta,\phi\right)g^{-3}h\\
 & +\sigma_{g}\left(\sigma_{h}h-1\right)u^{s_{r}*}\left(kgh,\alpha+\beta,\phi\right)u^{s_{p}*}\left(kg,\alpha,\phi\right)g^{4}h\\
 & +\sigma_{h}\left(1-\sigma_{g}g^{-1}\right)u^{s_{q}*}\left(kh^{-1},-\beta,\phi\right)u^{s_{r}*}\left(kh^{-1}g^{-1},-\alpha-\beta,\phi\right)h^{-6}g^{-3}\bigg]dgdhd\phi\\
 & =-\nu k^{2}u^{s_{k}}\left(\mathbf{k}\right)
\end{align*}

\bibliographystyle{aipnum4-2}
\bibliography{gurcan}

\end{document}